\documentclass[sigconf,nonacm,screen,authorversion]{acmart}

\usepackage[nolist]{acronym}
\begin{acronym}

    \acro{3DES}{Triple-DES}
    
    \acro{ACT-R}{Adaptive Control of Thought-Rational}
    \acro{AES}{Advanced Encryption Standard}
    \acro{ALU}{Arithmetic Logic Unit}
    \acro{ANOVA}{Analysis of Variance}
    \acroplural{ANOVA}[ANOVAs]{Analyses of Variance}
    \acro{API}{Application Programming Interface}
    \acro{AOI}{Area of Interest}
    \acroplural{AOI}[AOIs]{Areas of Interest}
    \acro{ARX}{Addition Rotation XOR}
    \acro{ATPG}{Automatic Test Pattern Generation}
    \acro{ASIC}{Application Specific Integrated Circuit}
    \acro{ASIP}{Application Specific Instruction-Set Processor}
    \acro{AS}{Active Serial}
    
    \acro{BDD}{Binary Decision Diagram}
    \acro{BGL}{Boost Graph Library}
    \acro{BNF}{Backus-Naur Form}
    \acro{BRAM}{Block-Ram}
    
    \acro{CBC}{Cipher Block Chaining}
    \acro{CFB}{Cipher Feedback Mode}
    \acro{CFG}{Control Flow Graph}
    \acro{CLB}{Configurable Logic Block}
    \acro{CLI}{Command Line Interface}
    \acro{COFF}{Common Object File Format}
    \acro{CPA}{Correlation Power Analysis}
    \acro{CPU}{Central Processing Unit}
    \acro{CRediT}{Contributor Roles Taxonomy}
    \acro{CRC}{Cyclic Redundancy Check}
    \acro{CTA}{Concurrent Think Aloud}
    \acro{CTR}{Counter}
    
    \acro{DC}{Direct Current}
    \acro{DES}{Data Encryption Standard}
    \acro{DFA}{Differential Frequency Analysis}
    \acro{DFT}{Discrete Fourier Transform}
    \acro{DIP}{Distinguishing Input Pattern}
    \acro{DLL}{Dynamic Link Library}
    \acro{DMA}{Direct Memory Access}
    \acro{DNF}{Disjunctive Normal Form}
    \acro{DPA}{Differential Power Analysis}
    \acro{DSO}{Digital Storage Oscilloscope}
    \acro{DSP}{Digital Signal Processing}
    \acro{DUT}{Design Under Test}
    
    \acro{ECB}{Electronic Code Book}
    \acro{ECC}{Elliptic Curve Cryptography}
    \acro{EEPROM}{Electrically Erasable Programmable Read-only Memory}
    \acro{EMA}{Electromagnetic Emanation}
    \acro{EM}{electro-magnetic}
    \acro{EMME}{Eye Movement Modeling Example}
    \acro{EU}{European Union}
    
    \acro{FFT}{Fast Fourier Transformation}
    \acro{FF}{Flip Flop}
    \acro{FI}{Fault Injection}
    \acro{FIR}{Finite Impulse Response}
    \acro{FPGA}{Field Programmable Gate Array}
    \acro{FSM}{Finite State Machine}
    
    \acro{GT}{Grounded Theory}
    \acro{GUI}{Graphical User Interface}
    
    \acro{HCI}{Human Computer Interaction}
    \acro{HDL}{Hardware Description Language}
    \acro{HD}{Hamming Distance}
    \acro{HF}{High Frequency}
	\acro{HRE}{Hardware Reverse Engineering}
    \acro{HSM}{Hardware Security Module}
    \acro{HW}{Hamming Weight}
    
    \acro{IC}{Integrated Circuit}
    \acro{IO}[I/O]{Input/Output}
    \acro{IOB}{Input Output Block}
    \acro{IoT}{Internet of Things}
    \acro{IRB}{Institutional Review Board}
    \acro{IP}{Intellectual Property}
    \acro{IRR}{Inter-Rater Reliability}
    \acro{ISA}{Instruction Set Architecture}
    \acro{ISCED}{International Standard Classification of Education}
    \acro{ITS}{Intelligent Tutoring Systems}
    \acro{IV}{Initialization Vector}
    
    \acro{JTAG}{Joint Test Action Group}
    
    \acro{KAT}{Known Answer Test}
    
    \acro{LFSR}{Linear Feedback Shift Register}
    \acro{LSB}{Least Significant Bit}
    \acro{LUT}{Look-up table}
    
    \acro{MAC}{Message Authentication Code}
    \acro{MAD}{Median Absolute Deviation}
    \acro{MIPS}{Microprocessor without Interlocked Pipeline Stages}
    \acro{MMIO}{Memory Mapped \acl{IO}}
    \acro{MSB}{Most Significant Bit}
    
    \acro{NASA}{National Aeronautics and Space Administration}
    \acro{NSA}{National Security Agency}
    \acro{NVM}{Non-Volatile Memory}
    
    \acro{OFB}{Output Feedback Mode}
    \acro{OISC}{One Instruction Set Computer}
    \acro{ORAM}{Oblivious Random Access Memory}
    \acro{OS}{Operating System}
    
    \acro{PAR}{Place-and-Route}
    \acro{PCB}{Printed Circuit Board}
    \acro{PC}{Personal Computer}
    \acro{PS}{Processing Speed}
	\acro{PR}{Perceptual Reasoning}
    \acro{PUF}{Physically Unclonable Function}
    
    \acro{RISC}{Reduced Instruction Set Computer}
    \acro{RNG}{Random Number Generator}
    \acro{ROBDD}{reduced ordered binary decision diagram}
    \acro{ROM}{Read-Only Memory}
    \acro{ROP}{Return-oriented Programming}
    \acro{RTA}{Retrospective Think Aloud}
    \acro{RTL}{Register-Transfer Level}
    
    \acro{SAT}{Boolean satisfiability}
    \acro{SEMOBS}{Self-Modifying Bitstreams}
    \acro{SCA}{Side-Channel Analysis}
    \acro{SHA}{Secure Hash Algorithm}
    \acro{SNR}{Signal-to-Noise Ratio}
    \acro{SPA}{Simple Power Analysis}
    \acro{SPI}{Serial Peripheral Interface Bus}
    \acro{SRAM}{Static Random Access Memory}
    \acro{SRE}{Software Reverse Engineering}

    \acro{TA}{Think Aloud}
	
	\acro{VLSI}{Very-Large-Scale Integration}
	
	\acro{WAIS-IV}{Wechsler Adult Intelligence Scale}
	\acro{WM}{Working Memory}
    
    \acro{TRNG}{True Random Number Generator}
    
    \acro{UART}{Universal Asynchronous Receiver Transmitter}
    \acro{UHF}{Ultra-High Frequency}
    \acro{UI}{User Interface}
    
    \acro{vFPGA}{virtual FPGA}
	\acro{VC}{Verbal Comprehension}
    
    \acro{WISC}{Writeable Instruction Set Computer}
    
    \acro{XDL}{Xilinx Description Language}
    \acro{XTS}{XEX-based Tweaked-codebook with ciphertext Stealing}
    
\end{acronym}

\usepackage{hyperref}
\usepackage{booktabs}
\usepackage{csquotes}
\usepackage{enumerate}
\usepackage{subcaption}
\usepackage{multirow}
\usepackage{rotating}
\usepackage{siunitx}
\usepackage{pdfpages}
\usepackage[inline]{enumitem}
\usepackage{xspace}
\usepackage{forest}
\usepackage{xcolor}

\usetikzlibrary{positioning}
\usetikzlibrary{backgrounds}

\definecolor{aoi_ui}{RGB}{1,115,178}
\definecolor{aoi_inout}{RGB}{222,143,5}
\definecolor{aoi_gate}{RGB}{2,158,115}

\definecolor{gate_irrelevant}{RGB}{254,177,88}
\definecolor{gate_relevant}{RGB}{44,102,140}

\definecolor{deleted}{RGB}{178,178,178}
\definecolor{inserted}{RGB}{14,130,27}

\newenvironment{inumerate}{
    \begin{enumerate*}[itemjoin={{, }}, itemjoin*={{, and }}, after={{.}}]
}{
    \end{enumerate*}
}

\newcommand{\ie}{i.\,e.}
\newcommand{\eg}{e.\,g.}

\newcommand{\etal}{et~al.\@\xspace}

\newcommand{\reversim}{\textsc{ReverSim}\xspace}

\AtBeginDocument{%
  }

\begin{document}

%\title[Combining Eye Tracking and Think Aloud to Describe Hardware Reverse Engineering]{Combining Eye Tracking and Think Aloud to Observe Human Factors in Hardware Reverse %Engineering -- A Mixed-Methods Approach\\
%Mixed Methods for Studying Human Factors in Hardware Reverse Engineering -- Eye Tracking and Think Aloud}

%\title{Eye Tracking and Think Aloud to Human Factors in Hardware Reverse Engineering -- A Mixed-Methods Approach}
\title[A Mixed-Methods Approach to Study Problem-Solving Processes in \ac{HRE}]{I see an \acs{IC}: A Mixed-Methods Approach to Study Human Problem-Solving Processes in Hardware Reverse Engineering}

\author{René Walendy} % R
\orcid{0000-0002-5378-3833}
%\email{rene.walendy@rub.de}
\affiliation{%
    \institution{Ruhr University Bochum}
    \city{Bochum}
    \country{Germany}
}
\additionalaffiliation{%
    \institution{Max Planck Institute for Security and Privacy}
    \city{Bochum}
    \country{Germany}
}

\author{Markus Weber} % M
\orcid{0000-0001-7775-807X}
%\email{markus.weber3@rub.de}
\affiliation{%
    \institution{Ruhr University Bochum}
    \city{Bochum}
    \country{Germany}
}

\author{Jingjie Li} % J
\orcid{0000-0001-6611-7496}
%\email{jingjie.li@ed.ac.uk}
\affiliation{%
    \institution{University of Edinburgh}
    \city{Edinburgh}
    \country{United Kingdom}
}

\author{Steffen Becker} % S
\orcid{0000-0001-7526-5597}
%\email{steffen.becker@rub.de}
\affiliation{%
    \institution{Ruhr University Bochum}
    \city{Bochum}
    \country{Germany}
}
\authornotemark[1] % hack in the second affiliation without causing duplicates

\author{Carina Wiesen} % C
\orcid{0000-0002-4403-1656}
%\email{carina.wiesen@rub.de}
\affiliation{%
    \institution{Ruhr University Bochum}
    \city{Bochum}
    \country{Germany}
}

\author{Malte Elson} % M
\orcid{0000-0001-7806-9583}
%\email{malte.elson@unibe.ch}
\affiliation{%
    \institution{University of Bern}
    \city{Bern}
    \country{Switzerland}
}

\author{Younghyun Kim} % Y
\orcid{0000-0002-5287-9235}
%\email{yhkim1@purdue.edu}
\affiliation{%
    \institution{University of Wisconsin--Madison}
    \city{Madison}
    \state{Wisconsin}
    \country{United States}
}

\author{Kassem Fawaz} % K
\orcid{0000-0002-4609-7691}
%\email{kfawaz@wisc.edu}
\affiliation{%
    \institution{University of Wisconsin--Madison}
    \city{Madison}
    \state{Wisconsin}
    \country{United States}
}

\author{Nikol Rummel} % N
\orcid{0000-0002-3187-5534}
%\email{nikol.rummel@rub.de}
\affiliation{%
    \institution{Ruhr University Bochum}
    \city{Bochum}
    \country{Germany}
}

\author{Christof Paar} % C
\orcid{0000-0001-8681-2277}
%\email{christof.paar@mpi-sp.org}
\affiliation{%
    \institution{Max Planck Institute for Security and Privacy}
    \city{Bochum}
    \country{Germany}
}

\renewcommand{\shortauthors}{Walendy et al.}

\begin{teaserfigure}
    \centering
    \includegraphics[width=0.7\linewidth]{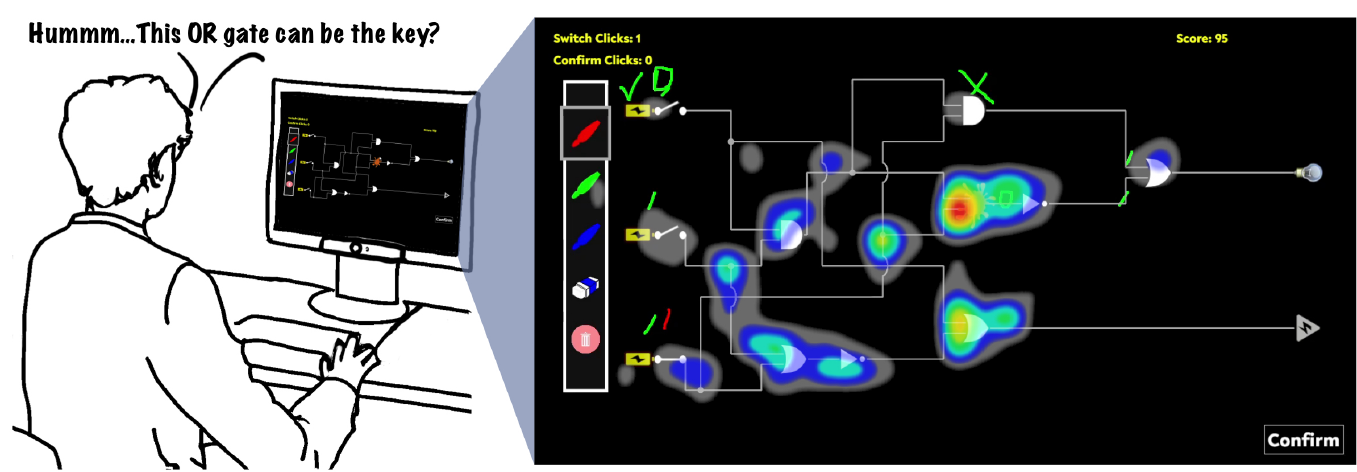}
    \caption{An illustration of a participant solving hardware reverse engineering tasks while thinking aloud, with eye tracking recorded, along with a heatmap of their visual attention (eye fixations) on the screen.}
    \label{fig:teaser}
    \Description{Figure 1 is an illustration of a participant, sitting in front of a computer, working on a study task. The computer screen displays a netlist of a small integrated circuit. An eye tracking device is attached to the bottom frame of the computer screen. The participant is using the computer mouse to interact with the task. The participant is thinking aloud: "Hah, this OR gate should be the key …" The right-hand side of the figure visualizes a heat map of the participants' eye-tracking data.}
\end{teaserfigure}
\begin{abstract}
%\textbf{Abstract}
%Problem
Trust in digital systems depends on secure hardware, often assured through \acf{HRE}.
%Contribution
This work develops methods for investigating human problem-solving processes in \ac{HRE}, an underexplored yet critical aspect.
%Approach
Since reverse engineers rely heavily  on visual information, eye tracking holds promise for studying their cognitive processes.
To gain further insights, we additionally employ verbal thought protocols during and immediately after \ac{HRE} tasks: Concurrent and Retrospective Think Aloud. %(\acs{CTA}, \acs{RTA}).
We evaluate the combination of eye tracking and \acl{TA} with 41 participants in an \ac{HRE} simulation.
%Findings
Eye tracking accurately identifies fixations on individual circuit elements and highlights critical components.
Based on two use cases, we demonstrate that eye tracking and \acl{TA} can complement each other to improve data quality.
%Impact
Our methodological insights can inform future studies in \ac{HRE}, a specific setting of human-computer interaction, and in other problem-solving settings involving misleading or missing information.
\end{abstract}

% keywords and CCS concepts



\maketitle

\section{Introduction}
\label{section:introduction}

With a strong reliance of society on information technology, people need to be able to trust an increasing variety of digital systems in their everyday lives. 
To build systems that can be truly trusted, we must be able to provide high assurance not only for all software layers but also for the underlying hardware.
At the core of such hardware, we find a multitude of microchips, commonly referred to as digital \acp{IC}.
To provide  assurance in hardware, analysts often employ \acf{HRE}, which is a crucial technique to detect, \eg, counterfeits~\cite{guin2014counterfeit}, intellectual property violations~\cite{subramanyan2014reverse}, or even malicious circuit manipulations~\cite{puschner2023red} in \acp{IC}.
\ac{HRE} consists of two fundamental steps~\cite{torrance2009state}:
First, analysts  reconstruct a blueprint of the thousands to millions of logic gates on the \ac{IC} -- the so-called \textit{netlist}.
Second, they analyze  this often very complex  netlist using a variety of methods tailored to their detection objectives.
Even though tools such as HAL~\cite{fyrbiak2019hal, wallat2019highway} exist, which partially assist human analysts during \ac{HRE}, success depends heavily on their experience, skills, and cognitive abilities for performing complex and diverse \ac{HRE} subprocesses~\cite{fyrbiak2017hardware, wiesen2021anatomy}.
Thus, similar to other application areas such as design engineering and medicine, \ac{HRE} represents a specific \ac{HCI} setting in which humans perform (computer-aided) complex problem solving using digital visualizations~\cite{Galantucci2006, le2010medical}.

When analysts navigate through a netlist, they have to integrate the current visual information with previously acquired knowledge about other parts of the netlist to make sense of its broader interconnections and functions.
We argue that  eye tracking is therefore highly suitable to gain a deeper understanding of how reverse engineers approach the analysis of a netlist.
 To interpret such eye tracking data holistically in terms of  participants' strategies, reasoning, insights, or reflections on errors, more information is needed.
A way to enrich eye tracking data with this kind of information is to let participants verbalize their thoughts as they solve an \ac{HRE} task, or to ask them to subsequently reflect on the task.
We propose that this combination of eye tracking and \ac{TA} is a promising and comprehensive approach to cognitive-factors research in \ac{HRE}.
% careful: Davies1995 has a major error in its evaluation, or Fig. 3 is flipped – there appear to be no errata published for this paper
However, evidence from similar fields, \eg, design engineering, marketing and cognitive psychology, show conflicting results as to whether verbalizations influence the behavior of the problem solver during a given task~\cite{Davies1995, Biehal1989} or not~\cite{Fleck2004, Ruckpaul2015}.
Therefore, it is not clear if and to what extent results from other studies can be transferred to \ac{HRE} problem-solving processes.
Hence, the primary objective of this work is to design, implement, and validate a mixed-methods approach that can subsequently be used to gain a comprehensive and nuanced understanding of the capabilities and skills influencing \ac{HRE} success.
Understanding how problem-solvers navigate in a netlist may inform the development of innovative hardware protection schemes.
It may further help tailor educational programs~\cite{wiesen2018teaching} on hardware security.

To investigate if and in which way eye tracking, verbal \ac{TA} protocols, and log files can be used to observe \ac{HRE} problem solving, we conduct an experiment with 41 participants.
Our main contributions are the following:
 \begin{itemize}
     \item We show that eye tracking provides high-resolution data on participants' visual attention  to individual circuit elements.
     \item We identify  \acl{TA} as suitable for gaining insight into reverse engineers' thought processes, with eye gaze--cued \ac{RTA} generating a higher quantity of codes, while \ac{CTA}  allows us to synchronize participants' verbalizations with their eye tracking data.
     \item We investigate the potential of  analyzing eye tracking and \ac{TA} data in combination
        and present two use cases:\begin{inumerate}
            \item mapping of \ac{TA} codes with eye tracking for the triangulation of participants' navigation patterns
            \item identifying  strategies used in an \ac{HRE} task with eye tracking, without needing to fully rely on \ac{TA}%
        \end{inumerate}
    \item Drawing on previous research on \ac{TA} and eye tracking, we advocate for methodological evaluations, propose enhanced combined analyses, and highlight the broader applicability of our approach to visually demanding problem-solving tasks in digital environments.
    In addition, our results provide insights for practitioners aiming at improving hardware security or developing educational tools pertinent to \ac{HRE}. 
 \end{itemize}

\section{Background and Related Work}
\label{section:background}

\subsection{Hardware and Netlist Reverse Engineering}
\label{subsection:background:hre}

In its broadest sense, reverse engineering describes the process of recovering the underlying specifications and design process of any man-made object by anyone else than the original designers~\cite{rekoff1985reverse}.
 \acf{HRE} in particular is often motivated by checking for the absence of malicious manipulations  also known as hardware Trojans, which may have been introduced by a contracted manufacturer~\cite{puschner2023red}.
Further applications include the identification  of \ac{IP} infringements such as counterfeit \acp{IC} which pose a major risk to the \ac{IC} supply chain~\cite{guin2014counterfeit}, or  failure analysis of \acp{IC} aimed at improving manufacturing processes~\cite{blythe1993layout}.

Azriel~\etal describe the reverse engineering of digital \acp{IC} as a two-step process~\cite{azriel2021survey}.
Initially, one obtains the circuit diagram of Boolean logic gates and memory elements and their interconnections -- the gate-level netlist -- from the \ac{IC}.
A single logic gate implements a small Boolean function such as NOT, AND, NAND or XOR.
While single gates are straightforward to grasp for humans, understanding the complex functions that emerge when several are combined quickly becomes a major challenge.
 Netlists can be  recovered using scanning electron microscopy of an actual \ac{IC}~\cite{torrance2009state} or, \eg,  by gaining direct access to design files describing the netlist~\cite{azriel2021survey}.
Reverse engineers then apply a wide range of methods for recovering and making sense of the higher-level structure and design rationale of the netlist at hand~\cite{meade2016netlist, subramanyan2014reverse, albartus2020dana}.
Given that this sense-making stage operates on the abstract digital description of a circuit, rather than the physical sample, it is often referred to as \textit{netlist} reverse engineering.
Analysts need to rely strongly on their own experience, technical skills, and cognitive abilities for manual analyses, as well as to adequately set up and apply methods from their semi-automated toolbox~\cite{fyrbiak2017hardware, becker2020exploratory, wiesen2021anatomy}.

State-of-the-art reverse engineering tools support analysts via interactive visualizations of the extracted circuit~\cite{wallat2019highway, thomas2021chipjuice, torrance2009state}.
Those visualizations appear to lend themselves well to the graph-based structure of electronic circuits and make them more accessible~\cite{wiesen2018teaching}.

\subsection{Prior Research on Problem Solving in \texorpdfstring{\ac{HRE}}{HRE}}
\label{subsection:background:related_methods}
While the technical steps in \ac{HRE} are often tedious to perform but generally well understood, little research has been done on the human problem-solving aspects.
Lee and Johnson-Laird performed an early laboratory study investigating how \ac{HRE} novices analyze simple Boolean circuits using fully manual approaches~\cite{lee2013theory}.
The authors define \ac{HRE} as a specific and poorly understood kind of human problem solving, requiring analysts to identify how each component influences the output of the circuit at hand, as well as how the different components depend on each other.

Later research applied the findings of Lee and Johnson-Laird to more complex, real-world \ac{HRE} settings~\cite{becker2020exploratory, wiesen2019promoting}, providing first valuable insights into the higher-level strategies and cognitive processes. 
 This strain of work led to a hierarchical model  that divides the \ac{HRE} process into \textit{reversing actions,} such as inspection and information gathering or strategy decisions, and \textit{source code development}~\cite{wiesen2021anatomy}.
A drawback of these initial investigations is that they are based on a complex training phase of human subjects and require massive evaluation efforts due to  manual annotation of log files.
This leads to small sample sizes and limits generalizability of these first studies on capabilities and skills in \ac{HRE}.

Recent efforts addressing these challenges have resulted in the development of \reversim~\cite{Becker2023Reversim}, a game-based simulation that mimics realistic \ac{HRE} subprocesses.
Notably, \reversim focuses on visually representing netlists, which are crucial for reverse engineers solving real-world \ac{HRE} tasks.
As \reversim enables fine control over these visualizations, it facilitates the consistent collection of eye tracking data, which we consider a promising basis for studying important \ac{HRE} subprocesses.
 At the same time, \reversim allows recording and quantifying interactions of reverse engineers, regardless of prior knowledge, in a standardized environment.

\subsection{Problem Solving and Eye Tracking}
\label{subsection:background:et}
Eye tracking measures a person's visual focus non-intrusively.
It estimates the positions of eye gazes from the captured eye images and the infrared reflections of pupils and corneas~\cite{majaranta-apc14}.
Eye tracking reveals rich and subtle dynamics of humans’ cognitive processes when they review materials, \eg, on a computer screen, without causing discomfort or requiring particular effort from  research participants.
Thus, researchers have been developing and  using eye tracking for decades to understand various psychological and physiological factors, including cognitive load~\cite{palinko-etra10}, personality traits~\cite{berkovsky-chi19}, health status~\cite{vidal-cc12}, and  problem solving~\cite{yoon-etra04,obaidellah-csur18}. 
Recently, using eye tracking to study engineering and computer-assisted tasks has received growing attention, especially in software engineering~\cite{obaidellah-csur18}. 
For example, researchers adopted eye tracking to investigate the individual differences in comprehending software programs~\cite{uwano-etra06}, which are related to different factors such as task familiarity~\cite{kather-toce21} and age groups~\cite{papavlasopoulou-idc17}. 
Sharafi~\etal leveraged eye tracking to identify people's problem-solving strategies when they manipulate data structures for programming~\cite{sharaf-tosem21}. 
Further, prior work utilized eye-tracking to reveal how people debug software programs, relating their performance differences to the problem-solving strategies applied~\cite{lin-te15}.
Beyond software engineering, previous research employed eye tracking in other tasks, particularly gamified tasks that involve significant visual interaction and navigation, to gain insights for educational, medical, and engineering applications~\cite{lee-idc21, lee-chb19}.
Recent studies have also proposed to use eye tracking in computer security research, including software reverse engineering~\cite{mantovani-sec22}.

 The eye tracking metrics used for studying problem-solving processes are primarily related to visual attention and its transition.
Fundamentally, eye tracking reports a time series of eye gaze positions.
The three most common abstractions of this time series are fixations, saccades, and scanpaths~\cite{sharafi-ese20, sharafi-15, sharafi-apsec15}.
Fixations are clusters of relatively stable eye gazes, standing for a basic unit of visual attention.
Saccades are rapidly moving eye gazes in between two consecutive fixations.
A scanpath is the resulting sequence of fixations due to the transition of visual attention.
Multiple spatial and temporal metrics are computed on the top of these abstractions, \eg, number of fixations, duration of fixations, and attention switching~\cite{sharafi-ese20, sharafi-15, sharafi-apsec15}.
Note that these metrics are often evaluated regarding the composition of visual stimuli, where researchers define their \acp{AOI}~\cite{salvucci-etra00}.
It enables them to analyze the fixation and saccade metrics within each  \ac{AOI} or across different ones~\cite{sharafi-ese20, sharafi-15, sharafi-apsec15}. 
Fixation metrics are more commonly adopted than saccades, as they retain richer information of cognitive processing~\cite{sharafi22,just-pr80}.
Prior research motivated us to employ eye tracking, especially fixation metrics, for HRE problem solving -- a task that is in particular visually demanding.
However, those eye tracking metrics alone may lack interpretability as a primary method to understand problem solving, though offering detailed measurements in its spatial context~\cite{cooke2005using}.
As such, prior work proposed to combine eye tracking with other study methods, \eg, \acf{CTA} or \acf{RTA} which elicits problem solvers' thoughts, to attain interpretable and fine-grained measurements at the same time~\cite{guan-chi06, Ruckpaul2015}.

\subsection{Concurrent and Retrospective Think Aloud}
\label{subsection:background:ta}
When conducting \ac{TA} in research studies, participants are asked to verbalize their thoughts to obtain information about their motivations, strategies, problems, or in general the \enquote{how} and \enquote{why} for a specific action. 
This method was first introduced in 1920 by Watson~\cite{Watson2009} with the goal of making thinking observable. 
Since then, \ac{TA} methods have been developed and evaluated in many domains and contexts.
In addition to \ac{CTA}, where participants are asked to verbalize their thoughts while solving a given task, \ac{RTA}, where participants are asked to verbalize the thoughts they had \textit{after} completing the task, has also been receiving attention in the research community~\cite{ericsson1993protocolanalysis}.
A common extension of \ac{RTA} is cued retrospective reporting, where participants are shown recordings of their problem-solving session, complemented by their eye movements and mouse operations.
Eye-gaze cueing has been shown to produce more comprehensive reports and improve participants' ability to recall their thoughts, even though the gaze cue can be distracting for some participants~\cite{elbabour2017eye}.
Both methods have individual advantages and limitations, and neither have been applied in the field of \ac{HRE}.
Therefore, one goal of our work is to  investigate both methods in conjunction with eye tracking during \ac{HRE} processes.

For \ac{CTA}, it is assumed that the verbalizations are mainly about actions and their outcomes~\cite{Taylor2000}, decision-making steps~\cite{Kuusela2000} or generally representations of short-term memory contents~\cite{ericsson1993protocolanalysis}.
We therefore assume a direct and unaltered access to participants' problem-solving approaches.
Also, immediate verbalizations are more synchronous to the eye movements than \ac{RTA}, where \ac{TA} is cued \textit{by} eye movement.
Of course, it is conceivable that \ac{CTA} might distract the participant from their task or otherwise affect their performance~\cite{ericsson1993protocolanalysis}.
% Careful here: the Ruckpaul paper cites Van Someren 1994 to claim that CTA can influence ET, but no mention of this is actually in the Van Someren work. There are other papers that follow Ruckpaul without checking. The following paper actually measures this:
Previous findings promote these assumptions at least in the field of design problem solving~\cite{Davies1995}.
Also, empirical studies have presented evidence that \ac{CTA} might skew eye tracking data~\cite{prokop20impact}.
However, a number of studies that did not find  influences on eye tracking data or problem-solving behavior \cite{Fleck2004, Lee2013strategic, Ruckpaul2015} challenge  those reports. 
Due to the inconsistent evidence, an investigation of \ac{CTA}'s influence on the \ac{HRE} process is part of the study's research questions.

Verbalizations generated by \ac{RTA} stem from both long-term and short-term memory~\cite{ericsson1993protocolanalysis} and are assumed to contain more statements on participants' final choices~\cite{Kuusela2000}. 
Cued retrospective reporting elicits even more action, \enquote{how} and metacognitive information than non-cued \ac{RTA}~\cite{vanGog2005}.
In addition, we can rule out the possibility that participants' performance during the task is influenced by \ac{RTA} and their performance can serve as baseline for comparison with the \ac{CTA} group.
For these reasons the cued \ac{RTA} method was chosen for the present study and its evaluation was included as part of the research questions.

\subsection{Research Questions}
\label{subsections:background:rqs}

From prior methodological work \textit{outside}  of \acf{HRE}, we derive that \acf{TA} and eye tracking may be promising techniques for investigating problem-solving processes involved in netlist reverse engineering.
Neither method has previously been used in the domain of \ac{HRE} problem solving.
Thus, it appears highly desirable to investigate their  usefulness in studying netlist reverse engineering behavior.
Given inconclusive findings in prior research, evaluating the potential interactions between eye tracking and \ac{TA} methods is essential for establishing a methodologically sound and robust experimental setup.
 To this end, we answer the following research questions:

 \begin{description}
     \item[RQ1] Can fixations obtained from eye tracking be used to observe behaviors within \ac{HRE} problem solving?
     \item[RQ2] How do \acl{CTA} and \acl{RTA} differ in revealing behaviors and approaches within \ac{HRE} problem solving?
     \item[RQ3] Does \acl{CTA} influence participants' performance, user experience, or eye movement?
     
 \end{description}

Based on the methodological insight from the three research questions above, we explore and discuss  a mixed-methods  design combining \ac{TA} and eye tracking.
We highlight the utility of both \ac{CTA} and \ac{RTA}, with a particular focus on the \textit{combined} analysis of verbal protocols and gaze behavior.
Thus, our overarching research question is:

\begin{description}
    \item[RQ4] How can eye tracking and \acl{TA} complement each other in describing \ac{HRE} problem solving?
\end{description}

\section{Methods}
\label{section:methods}
We conducted a lab study in which eye tracking was employed in combination with \ac{CTA} or gaze-cued \ac{RTA} while participants solved \ac{HRE} problems.
In the following sections, we provide details about the materials used, our participants, study procedures, as well as data collection and analysis.

\subsection{HRE Task Materials}
\label{subsection:methods:materials}

\subsubsection{\ac{HRE} Simulation}
\label{subsubsection:methods:materials:simulation}
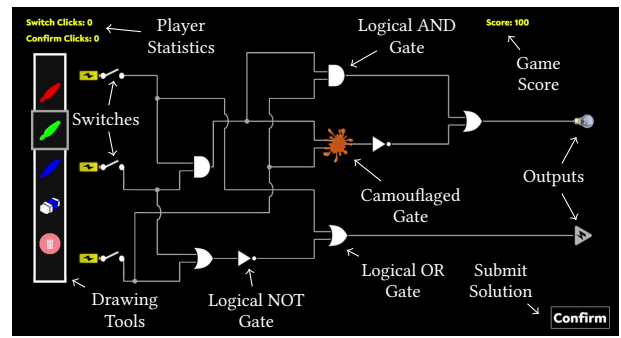
\begin{figure}[b]
    \centering
    \footnotesize
    \begin{tikzpicture}[]
        \tikzstyle{every node}=[white,align=center]
        \node (inputs) at (1.25,2.9) {Switches};
        \node (playerstats) at (2.25,4) {Player\\Statistics};
        \node (AND) at (5.25,4) {Logical AND\\Gate};
        \node (score) at (7,3.5) {Game\\Score};
        \node (camou) at (5.3,1.75) {Camouflaged\\Gate};
        \node (outputs) at (7.2,2.1) {Outputs};
        \node (submit) at (6.5,0.75) {Submit\\Solution};
        \node (OR) at (5.2,0.75) {Logical OR\\Gate};
        \node (NOT) at (3.25,0.35) {Logical NOT\\Gate};
        \node (drawing) at (1.5,0.35) {Drawing\\Tools};
        \draw[white]
            (inputs) edge[->] (1.35,3.4)
            (inputs) edge[->] (1.35,2.4)
            (playerstats) edge[->] (1.25,4.1)
            (AND)+(-0.4,-0.2) edge[->] (4.5,3.6)
            (score) edge[->] (6.6,4)
            (camou) edge[->] (4.55,2.35)
            (outputs) edge[->] (7.5, 2.65)
            (outputs) edge[->] (7.5, 1.55)
            (submit) edge[->] (7,0.3)
            (OR) edge[->] (4.45,1.15)
            (NOT) edge[->] (3.15,0.9)
            (drawing) edge[->] (0.8,0.75);
        \begin{scope}[on background layer]
            \node[inner sep=0pt, anchor=south west] (gamelevel) at (0,0)
                {\includegraphics[width=0.95\linewidth]{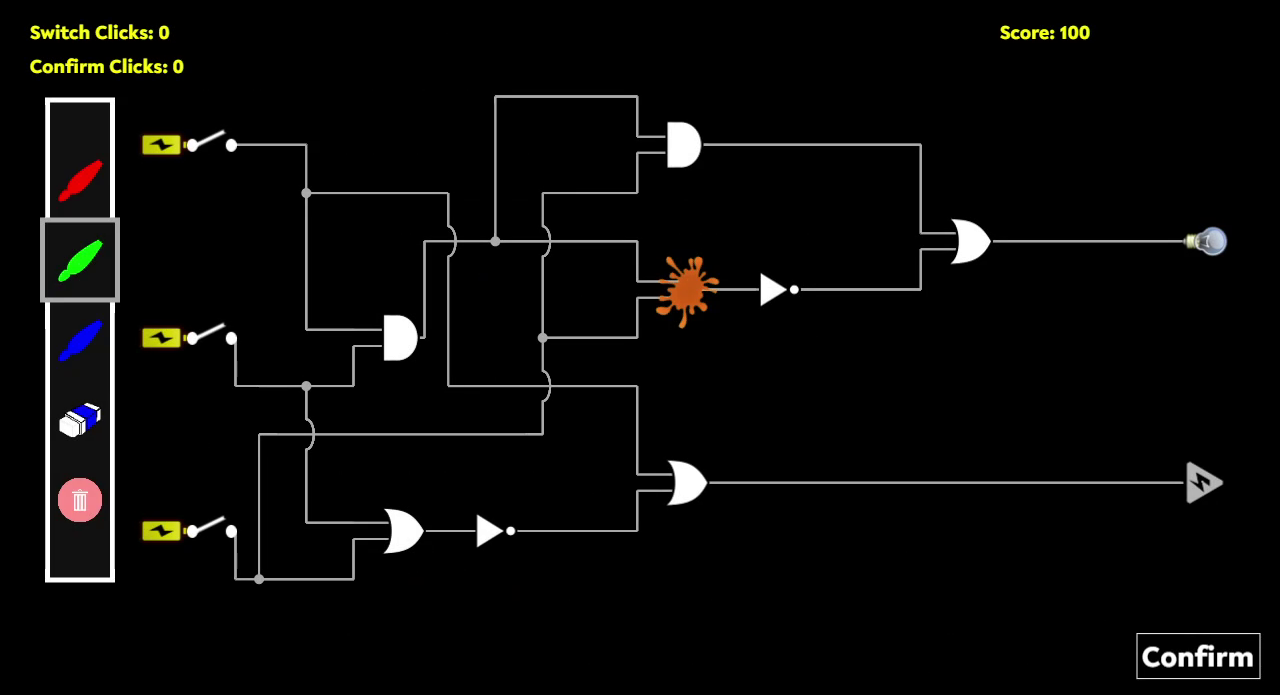}};
        \end{scope}
    \end{tikzpicture}
    \caption{An example level of \reversim. Participants need to understand the functionality of the circuit and then set the switches to the left such that the light bulb illuminates, whereas the danger sign must not be supplied with current. With the drawing tools they can annotate the circuit. The function of the gate in the form of an ink blot is hidden from the participants to make the solution of the level more difficult, simulating camouflaged gate obfuscation~\cite{cocchi2014circuit}.
    }
    \label{fig:methods:materials:simulation:level}
    \Description{Figure two is an example HRE task in ReverSim. It illustrates a small netlist of a circuit. On the left side of the circuit are three switches, labeled as “inputs”. On the right side of the circuit are a light bulb and a danger sign, which represent the goal state of the circuit and are labeled as “outputs”. Multiple wires which pass through different logic gates connect the switches to the light bulb and danger sign. The different logic gates have type-dependent shapes and are labeled according to their logic: “Logical AND gate” (semicircle), “logical OR gate” (arrowhead shape) and “logical NOT gate” (triangle). Furthermore, one of the logic gates is hidden by an orange ink blot. It is labeled “camouflaged gate”. The whole circuit is embedded in a user interface, which provides some further functions: On the bottom right is a confirm button, which is labeled with “submit solution”. On the left side “drawing tools” can be assessed to make annotations in the circuit. On the left top two counters indicate how many switch clicks and confirm clicks participants have made. It is labeled as “player statistics”. Last, on the right top participants' current “game score” is displayed.}   
\end{figure}

To administer the \ac{HRE} tasks central to the present work, we used \reversim, a computer game-based simulation of netlist reverse engineering problems~\cite{Becker2023Reversim}.
The simulation consists of multiple \textit{levels}, each representing one \ac{HRE} task.
Each task  comprises a Boolean circuit diagram that participants need to reverse engineer in order to advance to the next level.
Figure~\ref{fig:methods:materials:simulation:level} shows the user interface with an example circuit.
Specifically, in each task the participants need to reason about the functionality of the circuit in order to identify the binary input values that light all light bulbs  while \textit{not} triggering any danger signs.
To enter their solution, participants can interact with the three switches to the left.
Closing a switch powers the connected wire, corresponding to a binary input value of 1.
Each gate within the circuit takes the binary values on its input wires, applies its Boolean function, and generates a corresponding value on the output wire.
We illustrate this process using a straw man example in \autoref{fig:methods:materials:task:min_example}.
The effect of the chosen inputs is displayed by highlighting all powered wires once participants submit their solution.
Should the solution be incorrect, participants can start another attempt and revise the switch positions.
Participants can also annotate each circuit by using the mouse to draw onto the screen in three different colors.
\begin{figure}[b]
    \centering
    \includegraphics[width=0.95\linewidth]{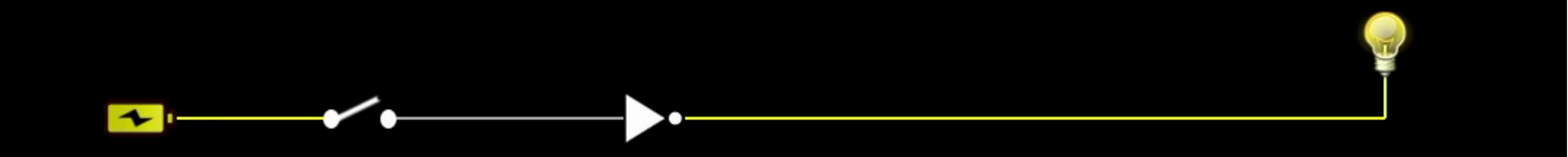}
    \caption{A straw man example of netlist reverse engineering. The goal is to light the bulb which needs a binary input value of 1. Knowing that the logical NOT gate can invert the signal from 0 to 1, we  make the switch open (0) to light the bulb.
    }
    \label{fig:methods:materials:task:min_example}
    \Description{Figure three shows a battery on the left side connected by a wire and two intermediate circuit components to a light bulb on the right side. The first intermediate component is an open switch, interrupting the wire. The wire then passes through a NOT gate before reaching the light bulb.}
\end{figure}
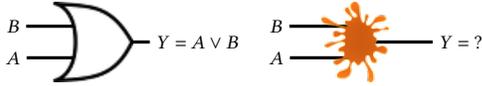
\begin{figure}[ht]
    \newcommand{\gatescale}{0.06}
    \centering
    \footnotesize
    \begin{tikzpicture}[scale=0.7]
        \node[inner sep=0pt,rotate=-90] (orgate) at (0,0)
            {\includegraphics[width=\gatescale\textwidth]{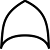}};
        \node (A) at (-1.5, -0.3) {$A$};
        \node (B) at (-1.5, 0.3) {$B$};
        \node (Y) at (2, 0) {$Y=A \lor B$};
        \node[inner sep=0pt,rotate=-90] (camou) at (5,0)
            {\includegraphics[width=\gatescale\textwidth]{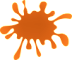}};
        \node (Ac) at (3.5, -0.3) {$A$};
        \node (Bc) at (3.5, 0.3) {$B$};
        \node (Yc) at (7, 0) {$Y=~?$};
        \begin{scope}[on background layer]
            \draw[very thick]
                (A.east) -- ([yshift=-0.3cm]orgate.center)
                (B.east) -- ([yshift=0.3cm]orgate.center)
                (Y.west) -- (orgate.center);
            \draw[very thick]
                (Ac.east) -- ([yshift=-0.3cm]camou.center)
                (Bc.east) -- ([yshift=0.3cm]camou.center)
                (Yc.west) -- (camou.center);
        \end{scope}
    \end{tikzpicture}
    \caption{Two gate symbols from the circuits used in the \ac{HRE} simulation. The left symbol depicts a standard logical OR gate with two inputs and a single output. The right symbol is specific to the \reversim environment and represents a camouflaged gate~\cite{cocchi2014circuit}. The logic function of this  circuit element is hidden from the participant, representing the case where the function of a gate could not be extracted from an \ac{IC} due to an obfuscation countermeasures.
    }
    \label{fig:obfuscated:gatetypes}
    \Description{Figure four shows two gate symbols from the circuits used in the HRE simulation. The left symbol depicts a standard logical OR gate with two inputs and a single output. The gate has the shape of an arrowhead. The top input is labeled with B, the bottom input is labeled with A. The single output is labeled with “Y = A or B”. The right symbol represents a camouflaged gate. The gate is obscured by an orange ink blot. The top input is labeled with B, the bottom input is labeled with A. The single output is labeled with “Y = question mark”.}
\end{figure}

\subsubsection{Netlist Reverse Engineering Tasks}
\label{subsubsection:methods:materials:tasks}
The tasks used in this work were taken from the \reversim level library. 
Each features three inputs and two outputs.
First, we selected four medium-complexity tasks, where each possible combination of light bulbs and danger signs appears exactly once, \ie, there were no two tasks with identical target output values.
Second, we included two tasks containing simulations of obfuscated circuit elements:
Covert gates aim at confusing reverse engineers by mimicking one type of gate when visually inspected, while actually implementing a different functionality~\cite{shakya2019covert}.
A camouflaged gate, on the other hand, is clearly identifiable as being obfuscated, but the actual functionality is hard to identify~\cite{cocchi2014circuit}.
\autoref{fig:obfuscated:gatetypes} shows an example of a camouflaged gate as a game element in \reversim,  visualized as an ink blot.

\subsection{Participants}
\label{subsection:methods:sampling}
In April and May 2022, we recruited 50 participants from a university in an English-speaking country.
Conditions for participation were a minimum age of 18 and sufficient English proficiency.
Each participant was compensated with 15 USD per hour.
We aimed at recruiting a diverse population regarding prior knowledge.
Therefore, we  not only advertised the study in electrical and computer engineering courses but also in other departments and on campus via email, flyers, and word of mouth.
The first nine participants were recruited to pilot the study.
Hence, we obtained a sample of 41 participants for the analyses reported below.
We randomly divided our participants into two groups with \ac{CTA} and \ac{RTA}, respectively.
20 participants were assigned to the \ac{CTA} group and 21 were assigned to the \ac{RTA} group.

\begin{table}[ht]
        \centering
        \caption{Basic demographics and prior knowledge score of our participants by assigned \ac{TA} condition.}
        \label{fig:methods:sampling:participants:demographics}
        \begin{tabular}{lp{0.18\textwidth}ccccc}
            \toprule
            \multicolumn{2}{l}{\textbf{Condition}}& \multicolumn{2}{c}{\textbf{CTA}} && \multicolumn{2}{c}{\textbf{RTA}}\\%
            && rel. & abs. && rel. & abs. \\
            %&& relative & absolute && relative & absolute \\
            \midrule
            \multicolumn{2}{l}{\textbf{Number of Participants}} && 20 &&& 21 \\
            \multicolumn{2}{l}{\textbf{Gender}} \\ 
            & male   & 60\% & 12 && 57\% & 12 \\
            & female & 40\% &  8 && 43\% &  9 \\
            \multicolumn{2}{l}{\textbf{Age Range}} \\ 
            & min && 18 &&& 18 \\
            & mean && 24.7 &&& 23.2 \\
            & max && 32 &&& 34 \\
            \multicolumn{2}{l}{\textbf{Education}} \\ 
            & secondary      & 40\% &  8 && 52\% & 11 \\
            & tertiary       & 60\% & 12 && 48\% & 10 \\
            \multicolumn{2}{l}{\textbf{Prior Knowledge Score}} \\ 
            & mean && 3.15 &&& 3.24  \\
            & SD && 1.27 &&& 0.83  \\
            \bottomrule
        \end{tabular}
        \Description{Table one contains basic demographics and the prior knowledge score of our participants for each TA condition. More precisely, this table shows 5 columns and 15 rows. In the first column the names of the categories are given, for example gender, age, and education. In columns two to five the relative and absolute frequencies for the CTA and RTA group are presented. Overall, the values indicate that groups did not differ regarding gender, age, education, and prior knowledge.}
    \end{table}
\begin{figure}[ht]
        \centering
        \includegraphics[width=0.85\linewidth]{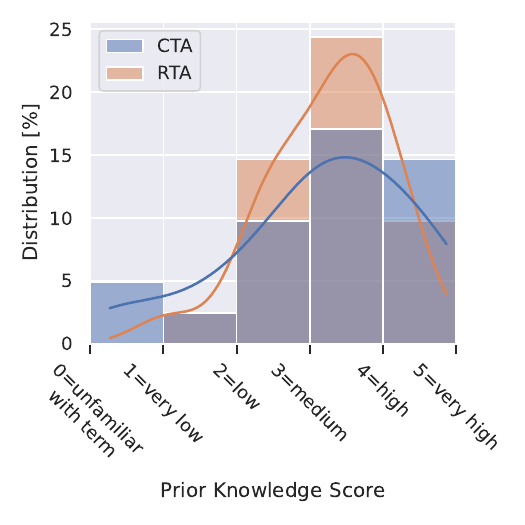}
        \caption{Distribution of prior-knowledge scores for both \ac{TA} conditions. Most participants self-rated their prior knowledge between \enquote{medium} and \enquote{high}.}
        \label{fig:methods:sampling:participants:pk}
        \Description{Figure five shows a histogram of the mean prior knowledge scores for both Think Aloud groups. The x-axis represents participants self-rated prior with five classes. Within each class the frequencies per group are represented by different colors. The y-axis shows the relative frequencies. The first class contains ratings between zero, which corresponds “not familiar with the term" and one, which corresponds to “very low”. All of them belong to the CTA group. The second class contains ratings between one and two, the later corresponding to "low". 2 percent belong to the CTA and 2 percent to the RTA group. The third class contains ratings between two and three, the later corresponding to “medium”. The fourth class contains ratings between three and four, the later corresponding to “high”. The fifth class contains ratings between four and five, the later corresponding to “very high”. Both groups a peak within the fourth class, containing also the reported mean value. A density curve added to the plot illustrates the homogeneity of the variances of both distributions.}
\end{figure}

After voluntarily agreeing on study participation, all participants answered a pre-study questionnaire on their age, gender and educational background including their majors of study.
\autoref{fig:methods:sampling:participants:demographics} shows a detailed breakdown of the demographics in our \ac{CTA} and \ac{RTA} groups.
In general, our sample  was young and educated.
28 of them had an educational background in electrical and computer engineering, while 13 had not. 
Furthermore, all participants self-rated their prior knowledge in 15 domains related to netlist reverse engineering.
The prior-knowledge scale was developed  alongside \reversim  and concerns areas such as Boolean algebra, digital circuits, as well as reverse engineering in general~\cite{Becker2023Reversim}.
It yields a combined score as the mean of the individual answers on a five-point Likert scale ranging from \enquote{Very Low} (1) to \enquote{Very High} (5).
All items offer a separate option \enquote{unfamiliar with the term}, which is represented as 0.
We report prior-knowledge distributions for both groups in~\autoref{fig:methods:sampling:participants:pk}, showing \enquote{Medium} as the most common score.

% technically the figure is now in the wrong section, but otherwise I don't get TeX to put it at the top of *this* page
\begin{figure*}[ht]
    \centering
    \includegraphics[width=\linewidth]{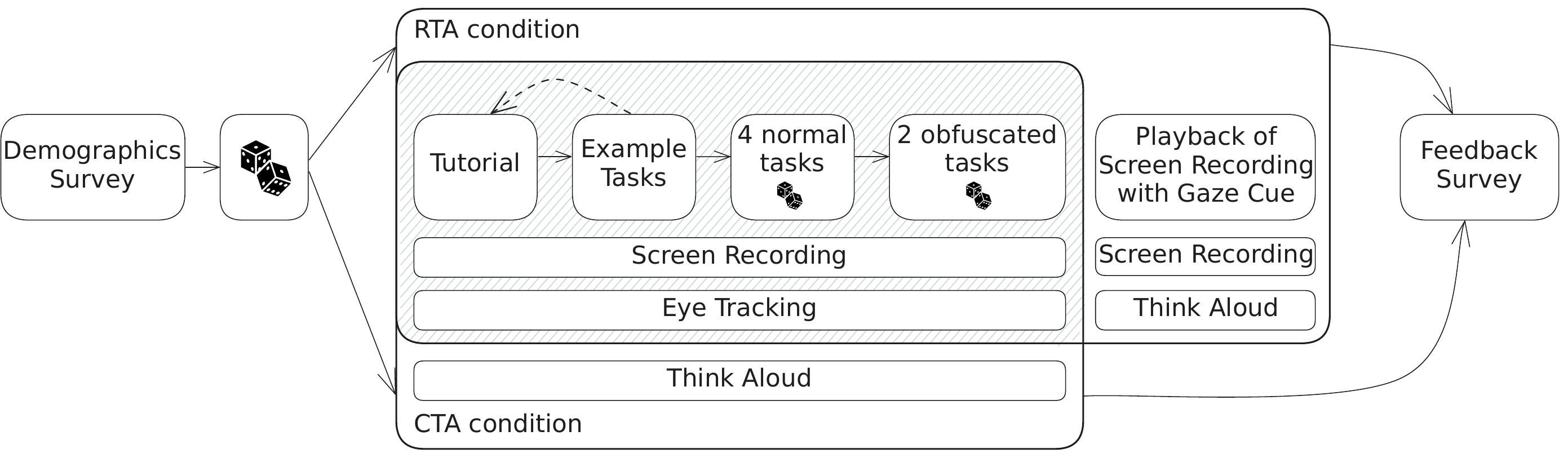}
    \caption{Overview of the study procedure. Participants first filled out a basic demographics and prior-knowledge survey and were then randomly assigned to the \ac{CTA} or \ac{RTA} condition. Both conditions contained the same set of tasks, during which screen captures and eye tracking data were recorded. Participants in the \ac{CTA} condition thought aloud while performing the tasks. Those in the \ac{RTA} condition thought aloud while reviewing a playback of their interactions with the tasks. Finally, all participants answered a feedback survey.}
    \label{fig:methods:design:flow}
    \Description{Figure six is a flow chart visualizing the study procedure described in Section 3.4.}
\end{figure*}

\subsection{Ethical Considerations}
\label{subsection:methods:ethics}

Our study was approved by our institution's \ac{IRB} and we secured participants' rights, safety and data privacy. 
Before participation, we informed participants about the study procedure, possible risks, and the right to withdraw at any time.
We minimized participants' fatigue with regular rest periods throughout the study. 
We recorded the computer screen, surveys, eye-tracking data and audio during think-aloud sessions.
Besides voice recordings, no personal information were collected.
We transcribed the audio recordings using an automated transcription service and manually curated them.
Through this process, we made sure no personal information is revealed in the transcripts for analysis and storage. 
Last, we stored all data on a secured department server. 

\subsection{Study Procedure}
\label{subsection:methods:design}
Accompanied by an experimenter, each participant completed the study procedure presented in~\autoref{fig:methods:design:flow}.
Participants gave voluntary informed consent, followed by a survey on  their background and demographics. 
Then, the experimenter walked participants through a tutorial about the basic concepts and interactions required to solve the tasks and gave instructions for thinking aloud.
Importantly, participants did \textit{not} receive explicit guidance on reverse engineering strategies.
To check whether participants understood the basic concepts and interactions, the experimenter instructed them to pass three minimal working examples of netlist reverse engineering tasks; otherwise, participants revisited the tutorial.
Participants in the \ac{CTA} group were instructed to practice thinking aloud while solving these example tasks. 

Next, we assigned the same six reverse engineering tasks described in \autoref{subsubsection:methods:materials:tasks} to both groups.
We first presented the  four tasks with medium complexity and then the two tasks with obfuscated gates, each in random order.
The experimenter instructed participants to rest between two consecutive tasks. 
Participants in the \ac{CTA} group thought aloud throughout the tasks. 
Conversely, participants in the \ac{RTA} group worked in silence and were guided to think aloud after solving all tasks.
When thinking aloud, these participants referred to a replay of their interactions with \reversim with their eye movement overlaid on the screen recording to support them in recalling their behaviors. 
In accordance with the recommendations by van Gog~\etal~\cite{vanGog2005}, participants were in control of the video playback.
After solving the tasks and thinking aloud, participants filled out a post-study survey about their experience with the tasks and the \ac{TA} method. 
Overall, each \ac{CTA} experiment took around 1 to 1.5 hours and each \ac{RTA} experiment consumed 1.5 to 2 hours.

\subsection{Data Collection}
\label{subsection:methods:data}
During the experiment, we collected logs and screen captures within the \reversim environment, as well as eye tracking and \ac{TA} recordings, in the manner presented below.
Furthermore, we collected participants' feedback after the study task.
We took particular care to ensure that all data is accurately time-synchronized, such as to allow cross referencing between the various types of data.

\subsubsection{Log Files}
\label{subsubsection:methods:data:logs}
Throughout the experiment, \reversim records all interactions, including drawing, interacting with switches, and submitting solutions.
All such events are stored in a time-stamped \textit{log file}, allowing us to precisely track participants' progress, solution time per level, and correctness of individual solutions.

\subsubsection{Eye Tracking Data}
\label{subsubsection:methods:data:et}
We used a Tobii Pro Nano eye tracker to collect eye tracking data with a sampling frequency of 60~Hz. 
We mounted the eye tracker and displayed the stimuli from the \ac{HRE} simulation on a 19-inch computer monitor with a resolution of 1280 $\times$ 1024 pixels. 
The monitor was placed  about 50 cm away from participants' seating position.
For each participant, we calibrated the eye tracker before collecting data, following a standard procedure~\cite{huang-dis23,zhang-chi19,hutt-hci22}.
We guided participants to take a seat and adjusted the eye tracker's tilt for them.
Next, we calibrated the system using the Tobii Pro Eye Tracker Manager software~\cite{TobiiTracker-23}.
The software displays nine white dots on different parts of the screen as targets for the participant to stare at.
It uses the calibration data to personalize the eye model for better tracking accuracy.

\subsubsection{\acl{TA} Protocols}
To ensure \ac{TA} quality, participants were asked to indicate their native language and to rate their proficiency in English. 
Based on their self-rating and our observations, all participants met general professional proficiency.
We captured their \ac{TA} verbalizations using a desk microphone and stored them in combination with a screen recording, such that the current state of the \ac{HRE} simulation as well as any mouse movements were available to provide context for later analysis.
Transcripts were automatically generated from the audio recordings for both \ac{CTA} and \ac{RTA} groups using Microsoft Office 365.

\subsubsection{Feedback Survey}
Following Ruckpaul~\etal~\cite{Ruckpaul2015}, we assessed whether participants were comfortable with both \ac{CTA} and \ac{RTA} settings.
All participants gave feedback on their personal experience of the experiment in three areas using an online questionnaire:
\begin{inumerate}%
    \item satisfaction with their personal task performance
    \item perceived difficulty of the task
    \item confidence when describing the tasks during \ac{TA}
    \item \ac{CTA} specific: how helpful it was to  think aloud to solve the tasks
    \item \ac{RTA} specific: how helpful the eye-gaze cued video playback was in remembering and describing what they were thinking%
\end{inumerate}%
All items across all areas consisted of a rating on a five-point Likert scale and an optional free-form answer field.

\subsection{Data Analysis}
\label{subsection:methods:analysis}
Below, we provide an overview of how we analyzed eye tracking and \ac{TA} data, respectively and jointly.

\subsubsection{Analysis of Eye Tracking Data}
To answer \textbf{RQ1}, our analysis of eye tracking data follows the steps below.

\paragraph{Data synchronization and cleaning.}
Before analyzing our data, we temporally aligned the data sources using timestamps attached to each recording.
We first synchronized the raw eye gaze data with the simulation log files recorded by \reversim.
This allowed us to subsequently extract the eye-tracking data for each \ac{HRE} task.
We removed data from individual \ac{HRE} tasks for a small number of participants due to poor quality or interruptions during the task.
For this purpose, we carefully cross-examined the logs from the eye tracker and the simulation server, screen recordings, and notes taken by the experimenter, resulting in the exclusion of 6 out of the 246 recorded tasks.

\paragraph{Fixation detection.}
To model participants' visual attention, we extracted their fixations from the raw eye gaze data. 
Our pipeline adopts the \texttt{I2MC}\footnote{\texttt{I2MC} is available under open-source license: \url{https://github.com/royhessels/I2MC/}} algorithm and applies Kalman filtering, interpolation of missing data, and 2-means clustering to detect fixations and address noise from the eye tracking data for our analysis~\cite{hessels-brm17}.

\begin{figure}
    \centering
    \includegraphics[width=0.95\linewidth]{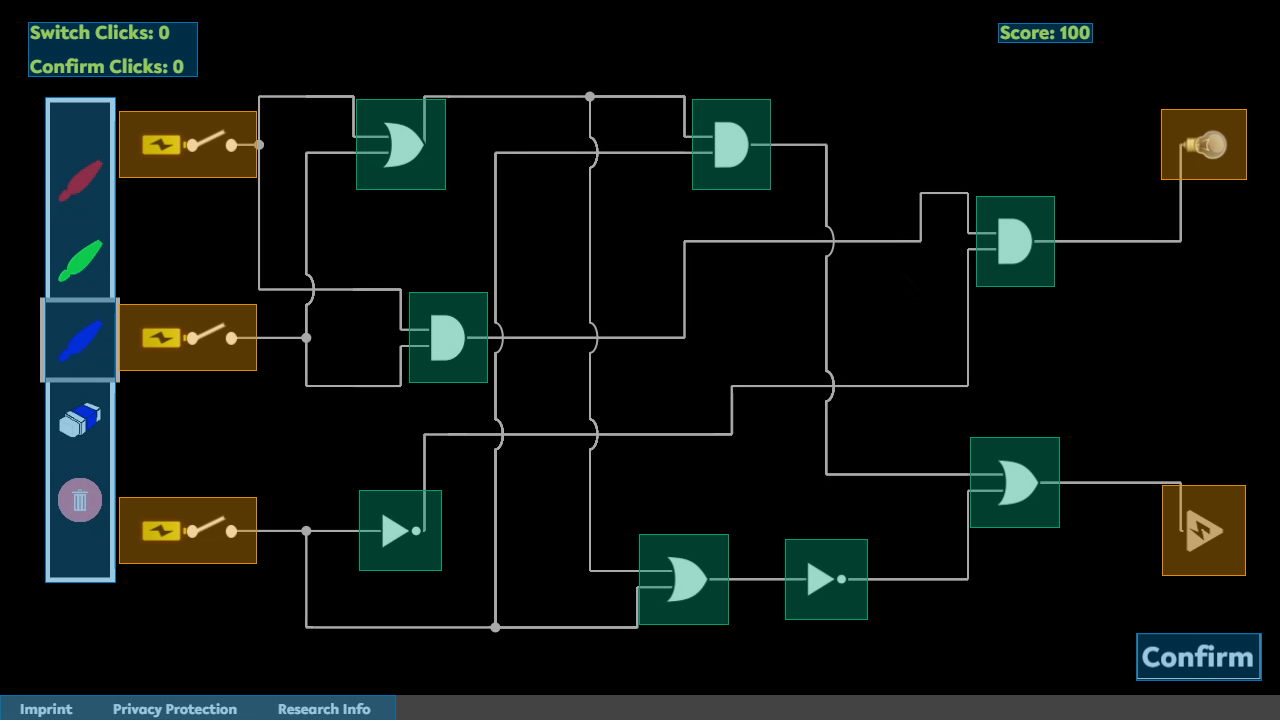}
    \caption{\acfp{AOI} defined in one of the \ac{HRE} tasks. The color indicates the \ac{AOI} category: \ac{UI} elements ({\color{aoi_ui}blue}), inputs and outputs ({\color{aoi_inout}orange}), and logic gates ({\color{aoi_gate}green}).}
    \label{fig:methods:analysis:aois}
    \Description{Figure seven shows a study task similar to figure two. Figure two contains a detailed description of a study task example. In the current figure, the areas of interest used for eye tracking are plotted and highlighted in different colors according to their AOI categories. Categories are “user interface elements” , “inputs and outputs” as well as “logic gates”.}
\end{figure}

\paragraph{Defining \acfp{AOI}.}
\ac{AOI} analysis enables us to evaluate fixation metrics with regard to individual visual elements of the \ac{HRE} simulation.
We defined \acp{AOI} on the screen for each of the six \ac{HRE} tasks.
Our \acp{AOI} were grouped into three categories: logic gate elements, circuit input/output, and \ac{UI} elements such as the controls for the drawing tools.
\autoref{fig:methods:analysis:aois} showcases the \acp{AOI} for one task.
Each \ac{AOI} covers a rectangular area around the center of each element.
We follow Goldberg and Helfman's guidelines to determine the granularity and sizes of \acp{AOI}~\cite{goldberg-beliv10}, despite there being no universal practice for it. 

\paragraph{Fixation metrics and analysis.}
We leveraged fixation metrics to evaluate participants' visual interest on \acp{AOI}.
We first measured \textit{fixation rate}, which is the ratio of the total number of fixations in each  \ac{AOI} to all \acp{AOI} of circuit elements~\cite{sharafi-apsec15}. 
A higher fixation rate suggests participants' greater visual interest  in an \ac{AOI}. 
Similarly, we computed the \textit{proportional fixation time}, which is the ratio of the sum of fixation durations in one  \ac{AOI} to the total duration time in all \acp{AOI} of circuit elements~\cite{bednarik-ijhcs12}.
In addition, we generated fixation heatmaps on images of the \ac{HRE} tasks to aid visual analysis.
In \textbf{RQ1}, we examine%
\begin{inumerate}%
    \item how \ac{HRE} tasks with different logic gate complexities affect participants' eye tracking metrics and
    \item the correlation of eye tracking features between different gates in relation to participants' problem-solving processes%
\end{inumerate}.

\subsubsection{Iterative Coding of Think Aloud Protocols}
\label{subsubsection:methods:analysis:coding}
We analyzed the \ac{TA} transcripts by qualitative coding  and content analysis~\cite{Krippendorff2019} to inductively develop a codebook in the light of \textbf{RQ2}.
One researcher checked and corrected the automatic transcriptions from both \ac{TA} data sets for errors.
The most common error sources were technical terms such as \textit{AND gate}, \textit{inverter} or \textit{wires}.
Simultaneously, the researcher applied content-based segmentation in order to prepare the transcripts for coding.
From the six recorded tasks for each participant, we selected both the first task they encountered and the task containing obfuscation (a camouflaged gate).
This way we wanted to capture the problem-solving behavior of the participants when they encountered a new task of each set for the first time.

Four coders individually applied an open coding procedure on those transcripts to categorize verbalizations.
We used screen recordings alongside to resolve verbal references to individual on-screen components.
All coders discussed the elicited codes and resolved disagreements.
To ensure that we accurately capture differences between \ac{CTA} and \ac{RTA}, the four coders generated a coding scheme for each separate \ac{TA} method before merging both codebooks.

Two coders individually applied the coding scheme on a training sample of 20\% of the transcripts and iteratively discussed differences in their coding in order to revise the codebook.
After this first phase \acf{IRR} reached $\kappa=.43$ (Cohen’s kappa).
We further refined the codebook to improve reliability by merging similar codes, further discussing disagreement and resolving ambiguity in code definitions.
Subsequent to this second  phase, the codebook was applied on 20\% of the sample to validate reliability.
Overall \ac{IRR} improved to $\kappa=.51$.
 However, specific codes  concerning participants' navigation behavior achieved an \ac{IRR} of $\kappa>=.71$.
Given that those codes are essential for mapping with the eye tracking data, we decided to apply the codebook with one coder and limited our analysis to those codes.
After creation of the final codebook (see  \autoref{section:appendix:codebook}), one of the involved coders deductively coded 48 tasks from 24 randomly chosen participants -- 12 from the \ac{CTA} and 12 from the \ac{RTA} group.
In our analysis and discussion, we report the frequencies of codes and compare the results from both \ac{TA} methods in relation to prior literature.
\subsubsection{Effects of the \ac{CTA} Method on Participants and Data}
As highlighted in \autoref{subsection:background:et}, asking participants to think aloud while solving a cognitively demanding task might affect their performance or impact eye tracking data.
In the light of \textbf{RQ3}, we therefore analyzed the effect of \ac{CTA} on our participants' experiences, as well as on their task performance and eye tracking data.
To verify that we  did not create a negative experience for participants by asking them to concurrently think aloud while solving the \ac{HRE} challenges, we analyzed their answers to the feedback survey.
In particular, we compared mean values of participants' self-reported confidence regarding the use of both \ac{TA} methods, as well as the perceived difficulty of the \ac{HRE} tasks.

Exercising caution regarding potentially detrimental effects of \ac{CTA} on the experiment itself, we then checked for differences between data acquired from the \ac{CTA} group compared to the \ac{RTA} group.
Here, the latter group serves as our non-\ac{CTA} control, given that those participants worked in silence.
Regarding task performance, we evaluated CTA's impact on participants' problem-solving time.
Furthermore, we counted the number of clicks on switches \textit{over par}, \ie,  the clicks that were unnecessary for an optimal solution, as well as the number of attempts required to solve each individual task and applied the same comparison.
In both cases, we used t~tests to determine whether the sample means differ significantly between the \ac{CTA} and \ac{RTA} groups.

Following the approach by Ruckpaul~\etal~\cite{Ruckpaul2015}, we further verified whether \ac{CTA} prolongs fixation duration compared to working in silence and thereby affects eye-tracking data.
For this purpose, we calculated the relative frequencies of fixation durations captured in either \ac{TA} conditions and compared the respective distributions.

\subsubsection{Use Cases for Combining Eye Tracking and \ac{TA}}
To  illustrate the potential of joint analysis of eye tracking and \ac{TA} data in the light of \textbf{RQ4}, we present two use cases studying how eye tracking features are correlated with data from \ac{TA} coding.

Our first use case concerns the level of individual code assignments.
Here we examined how fixation statistics, \eg, their position distributions, correspond to participants' behaviors observed in \ac{TA}.
 This approach relies on accurate time synchronization between eye gaze recordings and thought protocols.
While synchronicity can be manually achieved for \ac{RTA}, we focused our analysis on \ac{CTA} transcripts, where automatic synchronization is possible.
We extracted the start and end time for each occurrence of the relevant codes in the \ac{TA} transcript using the timestamps inserted by the automatic transcription service.
We then mapped the corresponding fragments of eye tracking data to each code occurrence and compared the fixation statistics between different codes.

 In our second, high-level use case, we compared eye tracking data from participants  who started an \ac{HRE} task with different strategies.
We first identified the initially applied strategy from the beginnings of \ac{RTA} and \ac{CTA} transcripts, using screen recordings for context where required.
After grouping participants by strategy, we then applied \ac{AOI} analysis to determine  mean fixation  durations for different \acp{AOI} and compared them across the groups.

\subsubsection{Statistical Calculations}
For statistical group comparisons, we used t~tests when normal distribution and homogeneity of variance were given.
We tested these with the Lilliefors~test~\cite{Lilliefors1967} and Levene's~test~\cite{levene1960contributions}.
If the assumption of normal distribution was violated we used Welch's~t~tests~\cite{WELCH1947}; if further homogeneity of variance was not given, Mann-Whitney~U~tests~\cite{Mann1947Test} were employed.
When multiple tests were calculated, Holm-Bonferroni~\cite{Holm1979Simple} correction was applied.
If variables were not metric, \eg, the code distribution between \ac{CTA} and \ac{RTA}, we used Pearson's  $\chi^2$~tests for comparison.
For the eye tracking data we also calculated Pearson's~$r$~\cite{Pearson1900}.
In general, we applied statistical calculations sparingly because\begin{inumerate}
    \item our sample was not very large, which is a risk for beta errors
    \item we wanted to prevent alpha error accumulation due to multiple testing%
\end{inumerate}

\subsection{Limitations}
We recognize several limitations in our study. 
First, our participant population is overall well-educated and young. 
Participants from other age groups or with other prior knowledge backgrounds may contribute to different observations in completing the \ac{HRE} tasks.
Second, though all participants met our requirements for English proficiency, it remains a question if speaking a non-native language affected their accuracy of \ac{TA} in this context. 
Also, it remains unclear whether verbalizations are representations of unbiased retrievals from memory during \ac{TA}. 
Nevertheless, future studies may use eye tracking to cross-validate the verbalizations during \ac{TA}.
Third, our work primarily analyzed participants' visual attention during \ac{HRE} problem-solving using fixations captured with eye tracking and \ac{AOI} analysis in line with prior work and guidelines~\cite{sharafi-15, obaidellah-csur18}.
We chose this particular focus because these features are most crucial to understanding visual attention, which is the foundation for solving \ac{HRE} tasks, and how to analyze them in combination with \ac{TA} was previously unknown.
Compared to fixations, the amount of information acquisition and cognitive processing during saccades is limited~\cite{sharafi22,just-pr80}.
Nevertheless, we encourage future studies to enable more kinds of analysis for \ac{HRE} from aspects other than attention, including stress and cognitive familiarity indicated by saccadic and pupil features respectively~\cite{skaramagkas-rbme21, kafkas-psycho15}.
In addition,  HRE problem solving may depend on other factors as well, such as the prior experience and working memory of the analyst~\cite{becker2020exploratory}. 
Quantifying the influences of these factors is not within the scope of this paper.
Despite these limitations due to the exploratory nature of our work, we believe our study's contribution is significant as the first work to design mixed methods for understanding \ac{HRE} problem-solving.

\section{Results and Discussion}
\label{section:results}

\subsection{RQ1: Analyzing \texorpdfstring{\acs{HRE}}{HRE} Problem Solving using Eye Tracking}
\label{section:results:rq1}

This section discusses how to interpret participants' processes to solve \ac{HRE} tasks from eye tracking features of individual circuit elements.

\begin{figure}[t]
    \centering
    \includegraphics[width=0.95\linewidth]{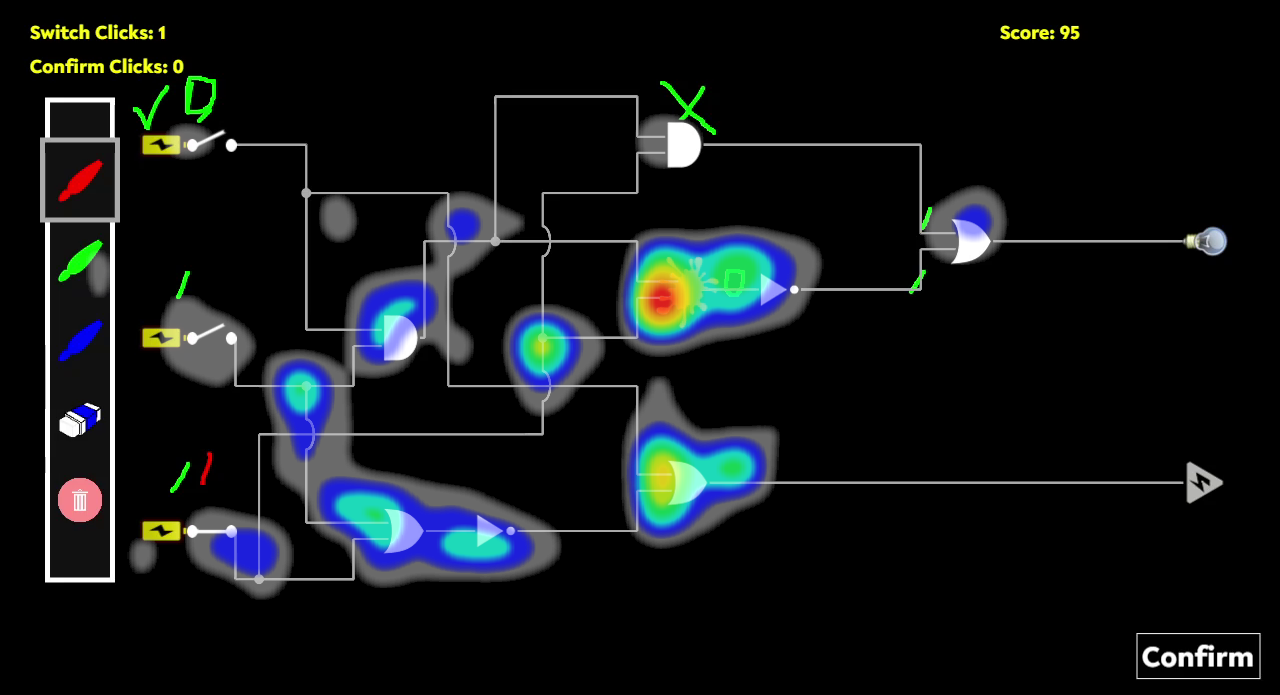}
    \caption{Heat map of raw eye gaze data for a single participant in an \ac{HRE} task involving a camouflaged gate, overlaid with the participant's annotations on the circuit. The camouflaged gate is represented by the orange ink blot in the center. It is evident that this participant focuses most of their attention on the camouflaged gate, while the input and output elements receive very limited attention.}
    \label{fig:results:rq1:heatmap_camou}
    \Description{Figure eight shows a study task similar to figure two. Figure two contains a detailed description of a study task example. In the current figure, a heat map is overlaid on the task. It reflects one participant’s raw eye gaze data captured while they solved the study task.}
\end{figure}

In \autoref{fig:results:rq1:heatmap_camou}, we present one participant's fixation heatmap as an example for visualizing their attention on different regions of an \ac{HRE} task.
This particular task features a camouflaged gate, which means that its actual logic function is obscured by an ink blot.
The example shows that the participant spent most fixations on the camouflaged gate, compared to other gates.
An insight gained from this is that even though uncovering the logic function of the camouflaged gate is not actually required to solve the \ac{HRE} task, the gate nevertheless presents a strong distractor.

While heatmaps are a helpful tool to visualize individual participants' behavior, statistical analyses on eye gaze behavior require an abstraction of \acp{AOI}:
In the following, we compute the number and duration of fixations on individual \acp{AOI}.
From this, we gain insight into the effect of gate types and obfuscation on eye gaze in \autoref{section:results:rq1:categories}.
In \autoref{section:results:rq1:correlation}, we investigate the effect of gate positions and their interconnection.

\subsubsection{Complexity of Gate Types}
\label{section:results:rq1:categories}

\begin{figure*}
    \centering
    \includegraphics[width=0.95\linewidth]{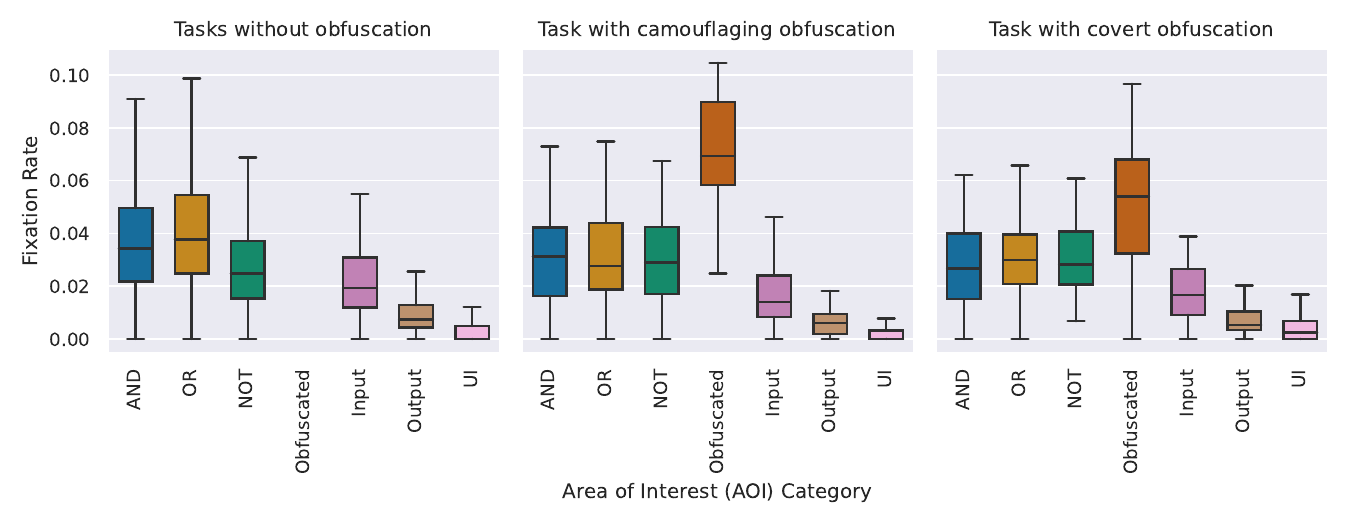}
    \caption{Statistics of fixation rate for each \ac{AOI} category under different task complexities (\ac{RTA} group). The basic logic gate types (AND, OR, NOT) receive similar attention across all types of the \ac{HRE} tasks, but are outweighed by both camouflaged and covert gates. Output and \ac{UI} elements receive little attention. In the statistics for proportional fixation time and the \ac{CTA} group, we attained very similar observations to this example.}
    \label{fig:results:rq1:gatetype_fixations}
    \Description{Figure nine shows three separate box plots, one for each HRE task type (regular, camouflaged, covert). The first plot contains six boxes, the second and third plot contain seven boxes, one box for each AOI type. It is evident that the obfuscated gate receives the most attention.}
\end{figure*}

We first demonstrate how tasks with and without obfuscated gates affect participants' visual attention.
\autoref{fig:results:rq1:gatetype_fixations} shows the statistics of fixation rates for each category of \acp{AOI} for the three different task types; \ie, without obfuscation, camouflaging obfuscation and covert obfuscation.
Overall, we observed that the fixation rates feature similar proportions between the \acp{AOI}  in  all three types of task.
 A comparison of proportional fixation times, shown in \autoref{section:appendix:fixations}, yields equal results.
For the tasks containing either type of obfuscated gate, those gates occupied a major fraction of visual attention compared to non-obfuscated ones.
This matches the initial observation from~\autoref{fig:results:rq1:heatmap_camou} where we concluded from the heat map that a considerable amount of attention is drawn to the camouflaged gate.

In addition, participants spent the least visual attention on \ac{UI} elements, despite the total size of this category being the largest.
Also, participants were more attentive to the input switches than the circuit's outputs.
This stands to reason because participants need to enter their solutions by clicking switches, while the outputs are static elements that require little reasoning.
Note that the variances are large in every type of task, revealing substantial individual differences between participants.

\subsubsection{Correlation Between Gates.}
\label{section:results:rq1:correlation}

\begin{figure}
    \begin{subfigure}[c][][b]{\linewidth}
        \centering
        \includegraphics[width=\linewidth]{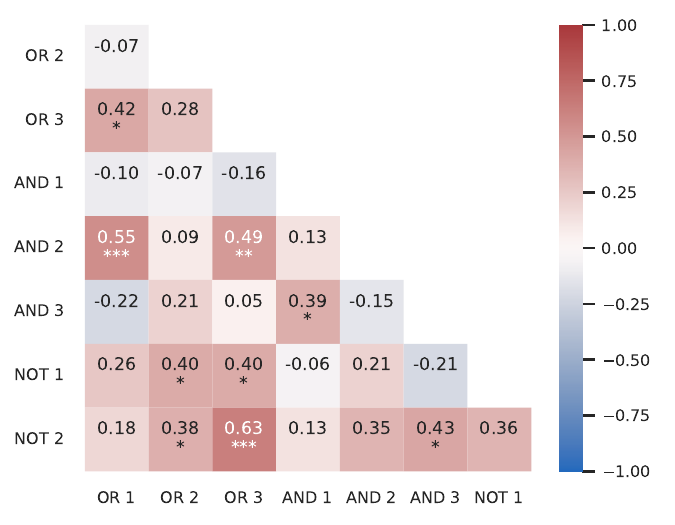}
        \caption{Correlation matrix. Asterisks in the matrix indicate significance levels: *:$p<0.1$; **:$p<0.05$; ***:$p<0.01$}
        \label{fig::results:rq1:corrmatrix:matrix}
    \end{subfigure}
    
    \bigskip
    \begin{subfigure}[c][][b]{\linewidth}
        \centering
        \includegraphics[width=0.65\linewidth]{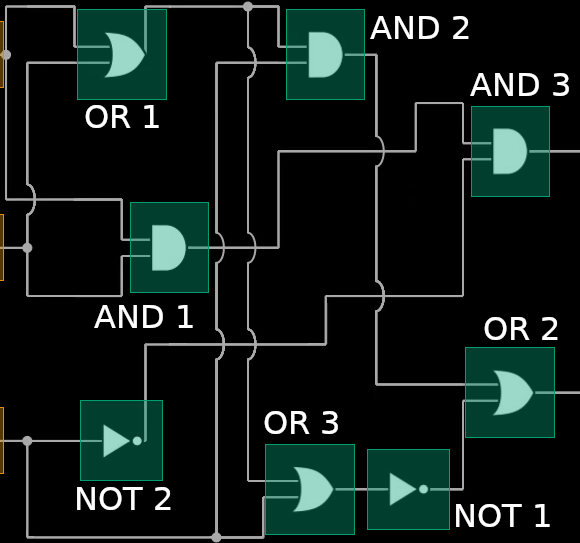}
        \caption{Condensed version of \autoref{fig:methods:analysis:aois} showing the corresponding \acs{AOI} in the underlying task.}
        \label{fig::results:rq1:corrmatrix:level}
    \end{subfigure}
    \caption{Correlation matrix between the fixation rates on all gate \acp{AOI} within one task. Significant correlations between fixation rates on \acp{AOI} are explained by their underlying gates being direct successors or having common inputs.}
    \label{fig::results:rq1:corrmatrix}
    \Description{Figure ten consists of (a) a correlation matrix and (b) a condensed version of Figure seven. Figure 10a shows positive and negative correlations between the areas of interest for all gates in the given task. None of the negative correlations are statistically significant. Positive correlations are more likely to be significant when the corresponding gates are connected to each other in the circuit. The highest positive correlation is about 0.63 whereas the highest negative correlation is -0.22. Figure 10b displays the gates and their wire connections. This figure can be used to compare the correlations of visual attention to the physical interconnections.}
\end{figure}

Beyond showing the influence of individual elements on problem solving, we employed eye tracking to investigate the underlying association between circuit elements. 
We computed the correlations of fixation rate and proportional fixation time between every pair of circuit element \ac{AOI} for all levels.
Specifically, \autoref{fig::results:rq1:corrmatrix} shows a correlation coefficient matrix for fixation rate in one task, across participants.
We find that the significant correlations visible in \autoref{fig::results:rq1:corrmatrix:matrix} correspond to gate pairs between which exists an immediate, or at least short, connection.

\subsubsection{Discussion of RQ1}
\label{section:results:rq1:discussion}
From the above results, we summarize the following takeaways. 
First, both heatmaps and \ac{AOI} analysis helped us identify where participants spent most visual attention during \ac{HRE} tasks. 
Notably, both analyses indicate that eye tracking on circuit diagrams as employed in the present setting is precise enough to resolve participants' focus on individual circuit elements.
From a problem-solving perspective, the distractive power of individual obfuscated gates is remarkable.
Even more so, camouflaged gates appear to draw participants' attention regardless of whether an understanding of their function is in fact required to solve the \ac{HRE} task at hand.
This insight may give rise to more efficient hardware protection schemes, which we will further discuss towards the end of this paper.

Second, we discovered correlations between fixation rates on different gates.
As these correlations correspond well with the interconnections of the gates within the netlist, we see this as a further indicator that eye tracking may be suitable to track participants' navigation within a netlist, for which we provide two use cases when answering \textbf{RQ4}.
While in the present example, we performed the required \ac{AOI} analyses after completion of the experiment, we consider it certainly feasible to do so in real time with dynamic \acp{AOI}.
The ability to perform such fast component-level \ac{AOI} analyses may, furthermore, open up a new avenue for integrating eye tracking into reverse engineering tools such as HAL~\cite{fyrbiak2019hal}.

\subsection{RQ2: Revealing Behaviors and Approaches through Participants' Think Aloud}
\label{section:results:rq2}
To further reveal participants' behaviors and approaches during \ac{HRE} problem solving, we analyzed their \ac{TA} transcripts using a coding technique based on qualitative content analysis as described in \autoref{subsubsection:methods:analysis:coding}.

\begin{figure}[t]
    \centering
    \includegraphics[width=\linewidth]{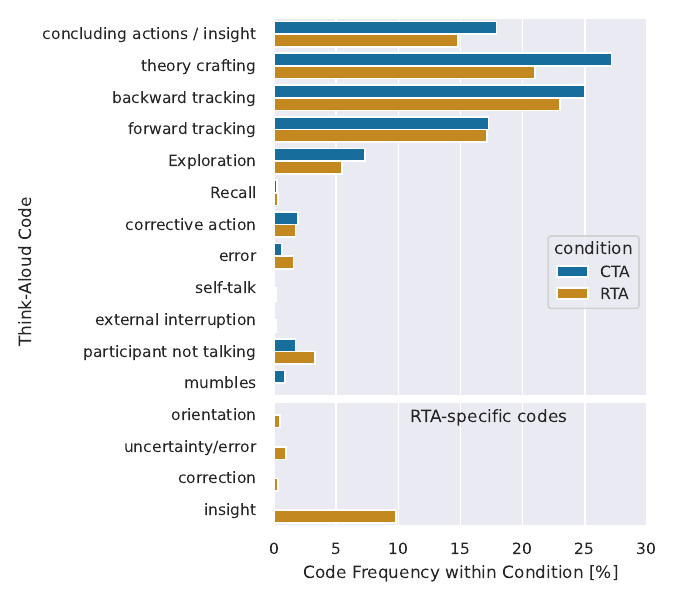}
    \caption{Relative  frequencies for each think-aloud code assignment compared between \ac{CTA} and \ac{RTA} conditions. Bar lengths indicate relative frequencies.  The bottom four codes are exclusive to the \ac{RTA} codebook. Either condition generates similar relative frequencies across all codes, even though, in absolute terms, \ac{RTA} generates more code assignments overall.}
    \label{fig:results:rq2:code_freq}
    \Description{Figure eleven is a bar chart showing the relative code frequencies for CTA and RTA groups, respectively. The most frequently assigned codes are “concluding actions / insight”, “theory crafting”, “backward tracking”, and “forward tracking” with between approximately 14 percent and 27 percent each. “Exploration” accounts for 6 percent and 8 percent of codes, respectively. Each of the other codes is infrequent with approximately 4 percent or less, except the RTA-specific code “insight” which accounts for 9 percent of all code assignments within this group. With exception of this code, relative frequencies are very similar between CTA and RTA groups.}
\end{figure}

\subsubsection{Final Codebook and Code Frequencies}
\label{section:results:rq2:codebook}
The final codebook consists of the  16 main codes shown in \autoref{fig:results:rq2:code_freq} with their observed relative frequencies.
Four of these codes were specific to the \acl{RTA} method.
The full codebook and code hierarchy including all sub-codes can be found in ~\autoref{section:appendix:codebook}.

For forward tracking we reached Cohen’s kappa of $\kappa=.71$ and for backward tracking $\kappa=.73$.
These codes were assigned to capture participants' navigation within the task, which is fundamental for understanding problem-solving processes in \ac{HRE}.
In particular, forward tracking was assigned when participants proceeded with a given input value from a switch or gate and tracked this signal further towards the next gate or the circuit's output; \ie, light bulb or danger sign.
Backward tracking was assigned when participants traced from the circuit's output or a gate towards the preceding gate or the switches.

We compared the relative  proportions between the codes common to both \ac{CTA} and \ac{RTA} and found no significant differences  (${\chi}^2 = 15.01, df = 15, p = .45$).
Interestingly, \ac{CTA} produces  an average of 38.7 codes while \ac{RTA} yields  45.9, which equates to approximately  19\% more codes in the \ac{RTA} condition.
Note that 11.5 percentage points of additional codes are allotted to the \ac{RTA}-specific codes, which describe behavior that occurred as participants recalled the task.
The code \textit{insight}, shown at the bottom of ~\autoref{fig:results:rq2:code_freq}, is by far the most prevalent of those codes and marks segments where participants came to new insights about the task or the quality of their solution while watching the video playback of their own actions.
This difference is largely explained by \ac{RTA} transcripts being significantly longer because participants were able to pause the video at will.
In contrast, the length of \ac{CTA} transcripts is limited by the duration of the \ac{HRE} tasks.
\ac{RTA} participants used the pause feature a total of 118 times, corresponding to 5-6 times per participant across all coded tasks.

We further found that rate of speech, measured in words per minute, differed only slightly between \ac{CTA} (94) and \ac{RTA} (98).
However, the presence of covert gates appears to influence rate of speech within the \ac{CTA} condition.
During the task with the covert gate participants verbalized approximately 24 words less per minute ($m=72.4, sd=30.5$) than in tasks without obfuscation ($m=96, sd=33.8$) or the task containing a camouflaged gate ($m=96.1, sd=37.4$).

\subsubsection{Discussion of RQ2}
\label{section:results:rq2:discussion}
The overall high complexity of \ac{HRE} tasks and participants' diverse strategies for solving them are reflected in our codebook.
Thus, for some codes, it is natural to allow ambiguity when coders are required to interpret participants' mental activities  from \ac{TA}.
 We observed from our iterative coding process that resolving this ambiguity is possible via extensive discussion of disagreements with at least two coders, however, doing so incurs a significant overhead.
In addition, the quality of codes applied to the \ac{RTA} transcripts may differ from those in \ac{CTA} due to some erroneous recollections during \ac{RTA} (\autoref{section:results:rq4:enhancing}).
Nevertheless, our current coding still reliably identified fundamental behaviors of \ac{HRE}, namely forward and backward tracking, from the \ac{TA} verbalizations.
When applying our codebook, coders should make a trade-off between saving efforts by reducing the codebook complexity and gaining more in-depth insights from intensive discussion.
In addition, both \ac{CTA} and \ac{RTA} appear to be appropriate \ac{TA} methods to study  \ac{HRE} problem-solving, considering the following trade-offs.
\ac{CTA} offers immediate verbalizations of participants' navigation and reasoning and allows high temporal synchronicity with eye tracking.
Nevertheless,  in very complex tasks, \ac{RTA} may offer more benefits than \ac{CTA}, as CTA participants might stop talking due to their strong focus on the task.
\subsection{RQ3: Differences Between \texorpdfstring{\acs{CTA}}{CTA} and \texorpdfstring{\acs{RTA}}{RTA} Groups}
\label{section:results:rq3}
To determine whether \ac{CTA} had an impact on participants' \ac{HRE}, we compared the \ac{CTA} and \ac{RTA} groups in terms of their performance, user experience, and eye-tracking data.
As the \ac{RTA} group did not think aloud while solving the levels but afterwards, their \ac{HRE} behavior served as a baseline for our comparison.

\subsubsection{Performance}
To measure performance of solving each of the six levels, we resort to three metrics: time, attempts, and switch clicks over par (see~\autoref{subsection:methods:analysis}).
\begin{figure}
    \centering
    \includegraphics[width=\linewidth]{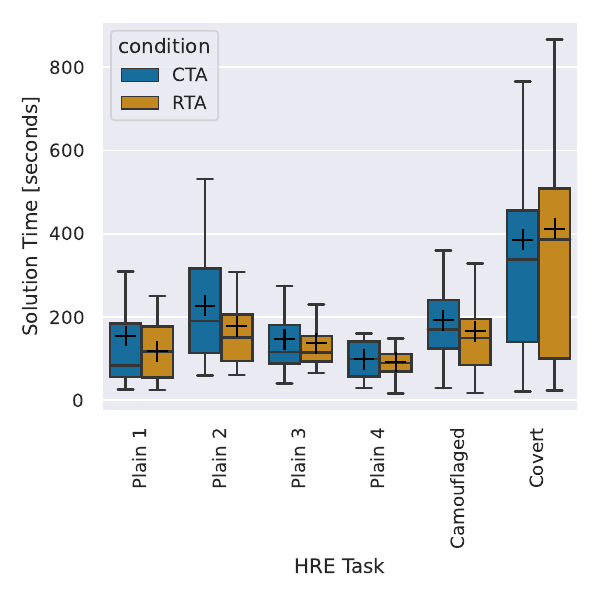}
    \caption{Distribution of participants' solution times for each \ac{HRE} task, divided by \ac{TA} condition. The four \enquote{Plain} tasks on the left do not feature obfuscation. An example task is given in \autoref{fig:methods:analysis:aois}. The two tasks on the right feature a single obfuscated gate. An example for the camouflaged gate task is given in \autoref{fig:methods:materials:simulation:level}. We observe similar solution times for both groups.}
    \label{subfig:results:rq3:solution_time}
    \Description{Figure twelve is a box plot with 12 boxes displaying the distribution of solution times per task of the CTA and RTA group, respectively. Solution times are similar between groups and levels, except for the task containing a covert gate. Here, for both groups, the mean solution time is twice as long as for the other tasks.}
\end{figure}
\autoref{subfig:results:rq3:solution_time} shows boxplots of the solution times per level for both groups.
To determine whether there were significant differences between the conditions, we calculated multiple Mann-Whitney-U-tests \cite{Mann1947Test} with Holm-Bonferroni correction \cite{Holm1979Simple}.
Even without correction, the groups did not differ significantly in their performance regarding time (ranges: $U = [201.0 - 259.0], p = [.10- .56]$).

Similarly, for switch clicks over par (ranges: $U = [126.0 - 205.0], p= [.16 - .85]$) and for number of attempts (ranges: $U = [138.5 - 192.5],  p= [.28 - .68]$) we observed no significant differences between \ac{CTA} and \ac{RTA} groups.

\subsubsection{Participants' Feedback on Task Difficulty}
In the feedback survey, we asked participants to rate the difficulty of the different task sets
on a 5-point Likert scale ranging from 1 (strongly disagree) and 5 (strongly agree).
As the performance of \ac{CTA} and \ac{RTA} participants did not differ significantly, we did not calculate inferential statistics for perceived task difficulty, but instead report mean values and standard deviations to reflect the participants' evaluations.

For the question, ``Solving the first four tasks~(puzzles) (\dots) was challenging to me.'', participants from the \ac{CTA} group ($m=2.45, sd=1.28$) as well as from the \ac{RTA} group ($m = 2.43, sd=1.25$) rather tended to disagree with mean answers between ``neither agree nor disagree'' and ``somewhat disagree.''

Conversely, for the question ``Solving the last two tasks (puzzles) (\dots) [with an obfuscated gate] was challenging to me.'', participants from the \ac{CTA} group ($m = 3.50, sd=1.15$) as well as from the \ac{RTA} group ($m = 3.86, sd= 0.91$)  rather tended to agree with means between ``neither agree nor disagree'' and ``somewhat agree.''

\subsubsection{Participants' Feedback on the \ac{TA} Procedure}

We also  report means and standard deviations of participant's self-rated confidence in describing the tasks by thinking aloud and participant's self-rated ease of verbalization for both groups, respectively.

For the question ``I felt confident when describing the tasks~(puzzles) during think aloud.'', participants from the \ac{CTA} group ($m= 3.95, sd=1.15$) as well as from the \ac{RTA} group ($m= 3.71, sd=1.23$) tended to ``somewhat agree.''

For the question, ``I found it easy to verbalize my thoughts.'', participants from the \ac{CTA} group ($m= 3.75, sd=1.33$) as well as from the \ac{RTA} group ($m= 3.57, sd=1.29$) also tended  to ``somewhat agree.''

Although the mean values of the Likert scale responses are very similar, the content of the responses differ between groups, when asked for explanation of their rating.
A participant in the \ac{CTA} group who ``strongly agreed'' regarding  their \ac{TA} confidence  argued \textit{``I felt more like I was explaining the logic aspects of the puzzle in my thoughts rather than random solutions so I felt comfortable with that, with an exception of the last task.''}
This  trade-off between difficulty and confidence was also mentioned by a participant who ``somewhat disagreed'': \textit{``I feel like I was too focused on trying to figure the puzzle out in my head than having to say it out loud. I got better at speaking when I knew what I understand. But when I get something wrong I would be confused and not know what to say.''}

Participants in the \ac{RTA} group expressed problems remembering their thoughts during the task;  \eg, \textit{``At some points I wasn't entirely sure what I was thinking,''} \textit{``(\dots)if this had been done during my solving of the problems, it would have gone much better (\dots)''}

The \ac{CTA} group was further asked to rate the statement ``I found it helpful to think aloud for solving the task (puzzle).'' With a mean of $m = 3.5,  sd = 1.43$ they scale between ``neither agree nor disagre'' and ``rather agree.''
One participant who ``strongly agreed'' mentioned: \textit{``Yes, thinking aloud and writing small notes helped me keep track of my strategy.''}
In contrast, for a participant who ``strongly disagreed'' it felt \textit{``(\dots) hard doing verbalizing and thinking at same time.''}

The \ac{RTA} group was asked to rate the statement ``I found it helpful to refer to the video playback with eye gaze cues when remembering and describing what I thought.'' With a mean of $m = 3.9,  sd = 1.22$ they ``rather agree.''
Nevertheless, participants  gave diverging explanations to this question in the free-form answer field.
One the one hand they stated that it was \textit{``definitely helpful because it reminded me of the way I approached the problem and what I started looking at and solving.''} but  on the other hand \textit{``(\dots) that it was more distracting than helpful.''}

\subsubsection{Eye Tracking Data}
\begin{figure}
    \centering
    \includegraphics[width=\linewidth]{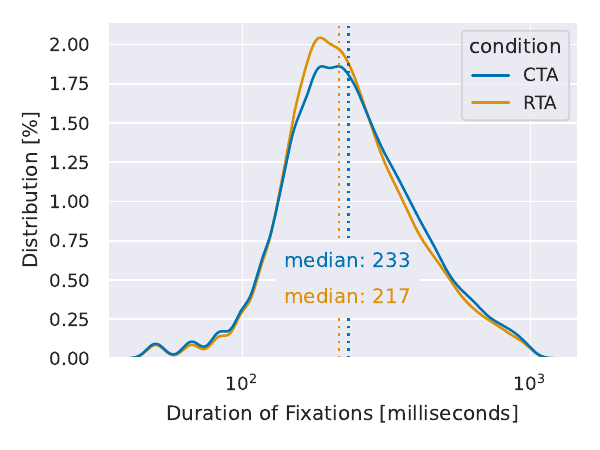}
    \caption{Distribution of fixation duration (kernel density estimation) for all participants in the \ac{CTA} and \ac{RTA} conditions (logarithmic scale). The two distributions are similar and approximately lognormal. The median fixation durations differ by about 16 milliseconds.}
    \label{section:results:rq3:fix_dur}
    \Description{Figure thirteen is a density plot of the fixation duration between CTA and RTA group. With a median duration of 233 and 217 milliseconds, respectively, it is evident that overall fixation lengths differ only marginally.}
\end{figure}

To identify whether the \ac{TA} method systematically influences eye gaze behavior, we compared relative frequency distributions of all fixations recorded within the \ac{CTA} and \ac{RTA} groups.
We find that both frequencies approximately follow a lognormal distribution as shown in~\autoref{section:results:rq3:fix_dur} and observe very similar shapes (\ac{CTA}: $mean=274\unit{\milli\second}$, $median=233\unit{\milli\second}$, $sd=158\unit{\milli\second}$; \ac{RTA}: $mean=267\unit{\milli\second}$, $median=217\unit{\milli\second}$, $sd=152\unit{\milli\second}$).
A Mann-Whitney U test shows that the observed 9-millisecond difference of mean values,  with \ac{CTA} exhibiting higher fixation lengths, is indeed strongly significant $(U=5.7 \cdot 10^8$, $p < 10^{-7})$.
This significance is probably due to the large sample of $n= 66,806$ individual fixations and should be interpreted with caution.
The median fixation duration shows a similar result with a difference of about 16 milliseconds.

\subsubsection{Discussion of RQ3}
\label{section:results:rq3:discussion}
From the above analysis we identify little impact of the \ac{CTA} method for all four tested effects.
Participants' performance in both the obfuscated and non-obfuscated tasks did not differ significantly between \ac{CTA} and working in silence, which is also reflected in their perceived task difficulty.

Regarding the effect on eye tracking, Ruckpaul~\etal observed mean \ac{CTA} fixations in their task that were about 55 milliseconds \textit{shorter} than in \ac{RTA}, however, did not obtain a statistically significant result ($p = 0.123$)~\cite{Ruckpaul2015}.
Contrary to those findings, our results indicate a slight positive and statistically significant difference in means.
We argue that, for the purpose of observing how visual attention is distributed on individual circuit components, this difference is of limited importance and comparability between \ac{CTA} and \ac{RTA} eye tracking data is generally given.
In summary, we have no evidence that \ac{CTA} systematically skews the data which our mixed-methods approach captures.

\subsection{RQ4: Eye Tracking and \texorpdfstring{\acs{TA}}{TA} as Complementary Research Methods}
\label{section:results:rq4}
By addressing RQ 1, 2 and 3 we showed that eye tracking and \ac{TA} are in isolation appropriate methods to investigate human problem solving in \ac{HRE}.
Combining eye tracking and \ac{TA} data might therefore be useful for achieving a more holistic explanation of \ac{HRE} behavior.
To answer the overarching research question on how the strengths of each method can be complemented, we present two use cases.
In \autoref{section:results:rq4:tracking_fixations}, we perform a  descriptive analysis localizing the prevalence of behaviors observed in different parts of the circuit, combining positional data obtained from eye tracking with time frames of individual behaviors observed in \ac{TA}.
In \autoref{section:results:rq4:startingpoints}, we show that, on a higher level of abstraction, participants' chosen starting points in the \ac{HRE} task leave characteristic patterns in eye tracking data.
Additionally, in \autoref{section:results:rq4:enhancing} we highlight how eye tracking can support \ac{RTA} in the form of eye gaze--cueing.

\subsubsection{Use Case I: Matching \ac{TA} Codes to Eye Tracking Fixations}
\label{section:results:rq4:tracking_fixations}

\begin{figure}[t]
    \centering
    \includegraphics[width=\linewidth]{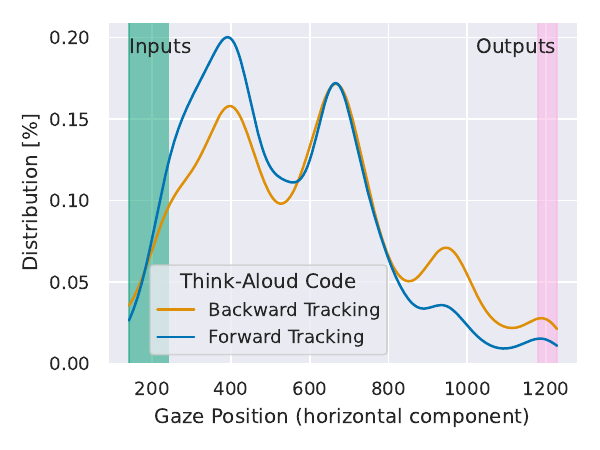}
    \caption{Distribution of the horizontal coordinate of fixations coded with forward and backward tracking actions (kernel density estimation), combined across all participants and all coded \ac{HRE} tasks. The shaded areas on the left and right show the position of the circuit's inputs, \ie switches, and the output symbols. During forward tracking, participants' visual focus is predominantly on the left side of the screen, towards the circuit's inputs and first set of gates. Backward tracking is more prevalent towards the right side of the screen, containing gates directly connected to either of the outputs. The three distinct peaks in both distributions are caused by the most prevalent horizontal position of gates across the different tasks.}
    \label{fig:section:results:rq4:trackingdirections}
    \Description{Figure fourteen displays multimodal distributions of eye tracking data for forward and backward tracking behaviors, respectively. The y-axis represents the density of the distribution in percent. The horizontal coordinate of the gaze position is shown on the x-axis. At the start of the x-axis, a green vertical bar represents the area in which circuit inputs are located. At the end of the x-axis, on the right, there is a pink bar that marks the circuit outputs.}
\end{figure}

An advantage of accurate time synchronization between eye tracking and \ac{CTA} protocols is the ability to precisely extract the eye tracking data for individual behaviors that were coded in the \ac{TA} protocols.

\paragraph{Setup.} In \autoref{section:results:rq2}, we have identified the codes \textit{forward tracking} and \textit{backward tracking} as frequent behaviors that we can detect in the \ac{TA} transcripts with high agreement.
However, localizing those behaviors is  extraordinarily tedious by manual coding of transcripts.
To demonstrate how eye tracking reflects this information,  we followed the steps below.  
We  first obtained two sets of fixation time series labeled \textit{forward tracking} and \textit{backward tracking}  using the timestamped code assignments from all \ac{TA} transcripts.
Each fixation represents visual attention to a specific coordinate on the screen.
For each of the two sets, we then calculated the distributions of the fixations along the horizontal axis of the screen. 
By comparing the resulting density plots,   we investigate  the assumption that \textit{forward tracking} occurs more often towards the inputs of the circuits, \ie, to the left, and that \textit{backward tracking} is prevalent towards the outputs on the right.

\paragraph{Results.}
\autoref{fig:section:results:rq4:trackingdirections} shows which locations on the circuit participants tended to fixate during episodes of forward or backward tracking.
From the plots it is apparent that \textit{forward tracking} is more prevalent at lower horizontal coordinates, \ie, towards the inputs of the circuit, whereas \textit{backward tracking} is more often occurring towards the outputs of the circuit.\footnote{Please note that each density plot is normalized, such that the area under both curves is 1.
This corrects for the fact that in total participants spent more time \textit{backward tracking} than \textit{forward tracking}, as \autoref{fig:results:rq2:code_freq} suggests.}
This result reflects our definition of forward and backward tracking behavior well.
We believe that with two coders and consensus after discussion, an even better fit of eye tracking and \ac{TA} could be reached.
Eye tracking data could further serve as an indicator for coding quality if only one coder is available.

\subsubsection{Use Case II: Identifying Strategies}
\label{section:results:rq4:startingpoints}
\begin{figure}
    \centering
    \includegraphics[width=0.95\linewidth]{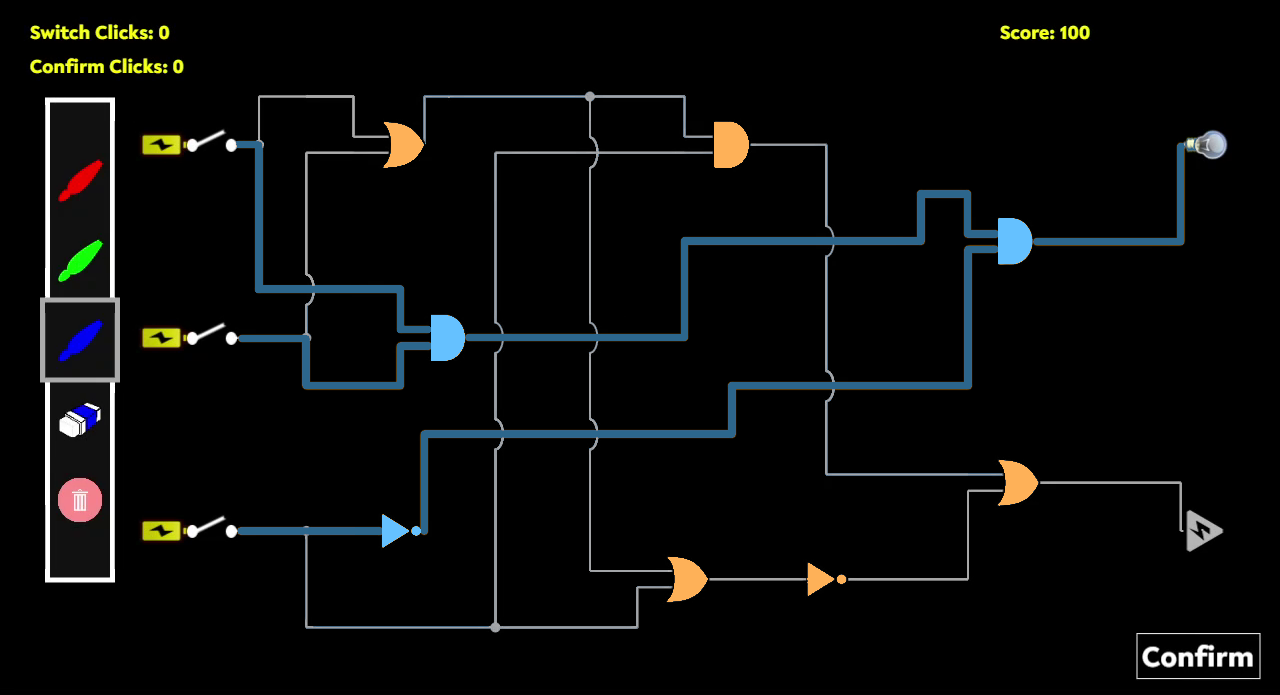}
    \caption{The structure of this \ac{HRE} task allows a shortcut: All three switch positions can be unambiguously determined by following the {\color{gate_relevant}blue} path backwards from the light bulb on the right. Inspecting the remaining gates colored {\color{gate_irrelevant}orange} is not required and does not yield additional information.}
    \label{fig:section:results:rq4:paths}
    \Description{Figure fifteen shows a study task similar to figure two. Figure two contains a detailed description of a study task example. In the current figure gates are colorized in blue or orange according to their importance for the task solution. The solution requires only blue gates to be inspected, while orange gates have no relevance. All blue gates feed, directly or indirectly, into a light bulb at the top right. Orange gates feed into a danger sign at the bottom right.}
\end{figure}
To find out whether eye tracking is suitable for identifying  task-specific  strategies in \ac{HRE} problem solving, we  explored individual participants' gaze behavior.

\paragraph{Setup.} First, we identified a sample task with a peculiar structure and formulated hypotheses for potential solution strategies:
It is sufficient to reverse engineer  the blue branch in \autoref{fig:section:results:rq4:paths} to unambiguously solve the task, \ie, all switch positions can be clearly determined by reverse engineering this one branch. 
Conversely, the  branch that ends in a danger sign alone does not provide sufficient information to solve the task.
Arguably, the best strategy  to solve this level is to apply \textit{backward tracking}  to  the blue branch from the light bulb, while the other branch  can be ignored.
Second, we grouped participants by strategy, using  the \ac{TA} transcripts to  identify at which of the two branches  they started reverse engineering.
If participants did not verbalize where they started, we extracted this information from the video.

Third, we used the eye tracking data to calculate fixation duration within the first 20 seconds during the task on the \acp{AOI} on both branches to capture how participants initially explore the circuit.
We then compared the average fixation duration of participants who started at the light bulb versus participants who started at the danger sign.

\paragraph{Results.} 
We found that participants who started reverse engineering at the light bulb spent an average of $1.44$ seconds on each \ac{AOI} on the path behind the bulb, while they fixated the other \acp{AOI} for $0.52$ seconds each.
Conversely, participants starting at the danger sign, spent an average of $0.99$ seconds on each \ac{AOI} on the light bulb path and $0.65$ seconds on each \ac{AOI} on the path feeding the danger sign.

In summary, the two strategies yield distinctive eye gaze patterns: Participants starting with the light bulb gazed more at the components of the blue branch in \autoref{fig:section:results:rq4:paths},  whether consciously or unconsciously.
Participants starting with the danger sign gazed less at the components of the blue branch and more at the  components of lesser importance.
We argue that one may extend this approach to generate eye gaze models for different \ac{HRE} problem-solving strategies identified from \ac{TA} within a small number of participants.
Using such models, automated analyses of eye tracking data could then enable the discovery of corresponding strategies within a large sample, where manual coding of full \ac{TA} protocols for the same purpose would be prohibitively time-consuming.

\subsubsection{ The Case for Gaze-Cued \ac{RTA}}
\label{section:results:rq4:enhancing}
During \ac{RTA}, participants sometimes had difficulty  remembering or verbalizing their problem-solving approaches during the task.
We coded these \ac{RTA} specific verbalizations as \ac{RTA}~orientation, \ac{RTA}~uncertainty/error, \ac{RTA}~correction, or \ac{RTA}~insight (see \autoref{section:results:rq2:codebook}).
This way, we could identify several instances in which participants reconsidered what they had just said because the captured eye tracking differed from their memories. 
To give an example from a participant who watched their video starting the first level: \textit{``(\dots) and I also tried to pay attention close to the beginning -- but I guess based on where my eyes were, that's not true -- where or what the end points were like.''}
The participant remembered that they first had to identify the goal state of the task, \ie~ light bulb or danger sign,  to subsequently start  reverse engineering from  there.
However, as this was their first task, they likely had not yet developed this strategy  and only used it in subsequent tasks.
During \ac{RTA}, the participant became aware of this mismatch of their memory and their actual behavior visible from their eye gaze recording.
This example indicates the potential improvement of \ac{RTA} by eye tracking.

\subsubsection{Discussion of RQ4}
\label{section:results:rq4:discussion}

In \textbf{RQ1}, we have shown that eye tracking has the precision to identify gaze behavior on individual gates through \ac{AOI} analysis.
However, participants’ actions are hard to interpret from eye tracking alone.
At the same time, \ac{TA} has shortcomings because it is often difficult to interpret which gates participants are applying an action to, even when researchers use screen recordings as context for coding.

Our two use cases demonstrate that combining both methods can be valuable for interpreting problem-solving behavior in \ac{HRE}.
First, eye tracking allows us to spatially locate specific behaviors identified from \ac{TA}.
Second, eye tracking can be used to differentiate between eye-gaze patterns resulting from different high-level reverse engineering strategies.
These techniques can reduce the amount of manual coding required in \ac{TA}, thus providing a means to increase the sample size that can realistically be analyzed.

An important consideration here is the choice of \ac{CTA} as the \ac{TA} method, as eye tracking data cannot be easily synchronized with \ac{RTA} protocols.
In addition, the \ac{CTA} method avoids the issue of participants not having a comprehensive memory of a task.
However, the \ac{RTA} method can be a valuable methodological adjunct when investigating challenging reverse engineering problems where participants are expected to become task saturated and therefore would be unable to talk while solving the task.

\section{Implications and Outlook}
\label{section:discussion}

In the following, we review the theoretical and practical relevance of our findings and outline future research directions.
First, we highlight the broader applicability of our methodological approach beyond \ac{HRE}.
We then discuss our findings in relation to prior research on \acl{TA} and eye tracking, and offer insight for improved combined analyses of both data sources. 
Second, our findings inform practitioners in enhancing hardware security or in developing educational resources for \ac{HRE}.
Finally, we suggest future research directions, including automated analyses to identify \ac{HRE} problem-solving strategies and further investigating cognitive factors relevant to \ac{HRE}.

\subsection{Methodological Implications}
\label{subsection:discussion:implications}

\paragraph{Reverse engineering as an \ac{HCI} phenomenon}
\ac{HRE} problem solving is not only visually demanding but also often involves navigating complex circuits. 
Lee and Johnson-Laird defined \ac{HRE} as ``the process of working out how to assemble components with known properties into a system that has the input–output relations of a target system,'' \ie, as a special kind of problem solving~\cite{lee2013theory}.
In particular, \ac{HRE} problem solving requires frequent  (re-)evaluation of assumptions as soon as a circuit's property -- \eg, an input -- changes.
This phenomenon is not exclusive to hardware security, as reverse engineering is a common problem-solving approach when people have to find and process information about a digital~\cite{canfora2007new} or physical system~\cite{Winnicki2017} that is not intuitively accessible (\eg, not well designed), not given (\eg, expert knowledge), lost due to poor documentation, or even intentionally hidden (\eg, through obfuscation or dark patterns).
Our methodological findings may therefore be used to inform studies in the above cases and may also be applied in various \ac{HCI} contexts.

\paragraph{Methodological considerations in relation to prior work}
We contextualize our methodological findings with previous work on methodological aspects of \ac{TA} and eye tracking.
Concerning \textbf{RQ2} (see \autoref{section:results:rq2}), we discovered that \ac{RTA} produces more overall codes but does not elicit significantly different information than \ac{CTA}.
Previous work has come to contradictory conclusions:
Some papers find that \ac{RTA}~\cite{Taylor2000, prokop20impact} generates a higher number of codes, while other papers find that \ac{CTA}~\cite{Kuusela2000, vanGog2005} produces a higher number of codes.
Prior research tends to agree that \ac{RTA} produces more insights into high-level reasoning and metacognitive reflections~\cite{Kuusela2000, Taylor2000, Ruckpaul2015, vanGog2005}\footnote{Van Gog \etal conclude that \textit{\ac{CTA}} produces more insights on reasoning but base their conclusion on absolute numbers of codes alone.
Taking the relative proportions into account, their findings appear to be generally in line with the other works.}, while participants in our study generated a substantial amount of high-level reflections during \ac{CTA} as well.
Second, our findings in \textbf{RQ3} (see \autoref{section:results:rq3}) evidence that \ac{CTA} does not affect task performance.
Prior work either agrees with our results~\cite{Fleck2004, prokop20impact} or reports either a small positive~\cite{Ruckpaul2015} or negative~\cite{Lee2013strategic} effect on performance.
Davies~\etal~\cite{Davies1995} further report a notable skew in problem-solving behavior when participants are asked to verbalize.
Considering a sample size of five participants per condition and an inconclusive result pertaining to the direction in which task behavior changes, this analysis should be interpreted cautiously.
In \textbf{RQ3}, we further observed that \ac{CTA} minimally prolongs eye fixations.
In contrast, prior work by Prokop~\etal~\cite{prokop20impact} finds that fixation duration is significantly \textit{shorter} when verbalizing.
Ruckpaul~\etal~\cite{Ruckpaul2015} generally agree, however, their observation does not reach statistical significance.
Prokop~\etal's work also suggests that the differing findings could be a result of how participants concentrate on the tasks during verbalization~\cite{prokop20impact}, which is open to further investigation.

In summary, we find inconsistent results in the literature to date.
We emphasize that the lack of concrete and widely applicable guidelines for combining \ac{TA} and eye tracking necessitates a basic methodological evaluation -- such as the one conducted in this paper -- to ensure a sound experiment design.

\paragraph{Suggestions for Mixed-Methods Designs}
In response to \textbf{RQ4}, we proposed two innovative semi-automated approaches to tightly combining eye tracking as a quantitative research method with qualitative content analysis of \ac{TA}.
Use Case I highlights how \ac{TA} codes can be used to select and compare specific episodes of eye gazes.
Use Case II suggests that eye tracking can help extract HRE problem-solving strategies more efficiently.
While researchers have previously applied eye tracking and \ac{TA} in the same experiment, some of this work uses eye tracking solely to provide a visual cue to participants during \acl{RTA}~\cite{elbabour2017eye, vanGog2005}.
Other work often evaluates the results from both methods separately~\cite{gegenfurtner2013transfer, salmeron2017scanning}.
Prior work that combines both sources either uses eye tracking for aligning AOIs or transcripts at a basic level~\cite{blascheck-etra16, guan-chi06}, filling silence periods in \ac{TA}~\cite{elling2012combining}, or 
providing additional qualitative context for manual \ac{TA} content analysis~\cite{cooke2005using}.
Our work expands on those approaches by integrating \ac{TA} protocols with quantitative and automated analysis of eye gaze.
We suggest that our joint analysis may offer more fine-grained insight into problem-solving behavior with reduced analysis effort, without extensive changes to existing experiment designs.

\subsection{Practical Applications}
\label{subsection:discussion:applications}

\paragraph{Enhancing hardware protection by cognitive obfuscation.}
A better understanding of \ac{HRE} problem-solving processes may help to protect \acp{IC} against adversarial \ac{HRE}, \eg, by competitors or hostile nation-state actors.
While traditional obfuscation aims at defeating reverse engineering tools and algorithms~\cite{zhang-tvlsi15}, recent work introduced the concept of ``cognitive obfuscation''~\cite{wiesen2019towards}.
This twist on the \ac{HCI} framework attempts to hamper \textit{human} understanding of a circuit.
We observed that camouflaged and covert gates can draw considerable attention.
Hardware designers may thus use those traditional obfuscated gates to introduce a false lead into an attacker's problem-solving process:
By selectively obfuscating parts of the circuit unrelated to the security-critical areas, they may shift attackers' attention away from the relevant components.
With this selective obfuscation, defenders can use obfuscated gates sparingly and thus economically while still wasting the attackers' time and resources.
The eye tracking metrics introduced in our research could further be used to construct models that quantify the efficacy of such obfuscation.

\paragraph{Improving education in hardware security.}
Recent massive investments~\cite{uschips2022,euchips2022} in domestic semiconductor fabrication are creating a major demand for the training of new talents in hardware security.
In the field of \ac{HRE}, where experts are already scarce, a serious shortage in hardware security educators thus arises.
This demand for education may be supplemented with computer-aided tools for independent learning.
Our method will motivate the design of learning content in a tutoring system for \ac{HRE} novices. Specifically, a promising method is the application of \acfp{EMME}:
Using eye tracking and \ac{TA}, one captures experts' explanations as well as visualizations of their gaze locations as they solve an \ac{HRE} problem.
\acp{EMME} foster learning by guiding attention, illustrating advanced perceptual strategies, and inducing a stronger social learning situation as learners watch experts performing a task~\cite{Krebs2019, Krebs2021}.
We consider \acp{EMME} to be well-applicable to \ac{HRE} education, following our evidence from \autoref{section:results:rq4:startingpoints} that differing netlist analysis strategies are indeed reflected in the corresponding eye gaze recordings.

\subsection{Future Research Directions}
Our work suggests multiple research directions towards a more comprehensive and less complex analysis of the human aspects of \ac{HRE}, which can lead to more trusted hardware devices.
We identify  two major directions in the following.

\paragraph{Automating analysis of \ac{HRE} strategies using eye tracking and \ac{TA}}
\autoref{section:results:rq4:startingpoints} demonstrates that our participants' problem-solving strategies varied. 
We expect that manually identifying problem-solving strategies will become less feasible as the size and complexity of hardware circuits grow. 
Consequently, we suggest in future research to automate this process based on the following aspects.
First, future approaches can further automate \ac{AOI} definition, especially for dynamic visual stimuli (\eg, pop-up notices) in real-world HRE scenarios, which can be more challenging. 
Prior research in software engineering has shown that it is possible to define dynamic \acp{AOI} automatically from fixation saliency during software code navigation tasks~\cite{sharafi22}.
Second, segmentation and labeling of eye tracking data can be streamlined with the aid of machine learning models tuned for coding and labeling the \ac{TA} data, which will then enable more fine-grained analysis of the temporal phases in problem solving~\cite{deane-brm23}.

\paragraph{Studying the influences of  cognitive factors within \ac{HRE} problem solving.}
This work is exploratory with respect to understanding \ac{HRE} processes based on participants' visual attention. \ac{HRE} problem solving entails multiple sub-processes (see \autoref{section:results:rq2}), which are influenced by different cognitive factors, such as prior knowledge and working memory~\cite{wiesen2021anatomy}.
These factors might be behind the varieties in problem solving  that we have observed (see \autoref{section:results:rq4}), which encourages a more granular analysis of eye-tracking data for these sub-processes and factors. 
Our case study in \autoref{section:results:rq4:tracking_fixations} exemplifies such potential, as fine-grained behaviors can be interpreted from just a few seconds of eye tracking data.
By recruiting more participants and introducing additional study instruments, future studies could extend our methodology to quantifying the effect of different cognitive factors, such as working memory~\cite{baddeley1992working}, in these sub-processes.
Further, additional physiological data modalities could be analyzed jointly with eye movement, including pupil dilation, electroencephalogram, etc. Prior work indicates the benefits of such multi-modal data for understanding of human processes during screen interaction, \eg, with respect to spatial abilities~\cite{slanzi-if17,sharaf-tosem21}.

\subsection{Conclusion}
\label{section:conclusion}
Understanding the human aspects of \acf{HRE} is a crucial step  in building more secure hardware. 
However, an in-depth understanding  remains challenging through traditional methods, \eg, log file analysis, due to the complex problem-solving process of \ac{HRE}. 
Recognizing visual processing as a key in people’s \ac{HRE} problem-solving process, we  contribute an innovative mixed-methods  \ac{HRE} study that  combines eye tracking and \acf{TA} to gain  deeper insight into such processes. 
Based on our study with 41 participants, we  gathered evidence in support of the hypothesis that fixations are an appropriate eye-tracking metric to describe participants' visual attention within an \ac{HRE} task.
In particular, \acf{AOI} analysis with fixation rates is valuable for quantifying attention to individual circuit elements.
Furthermore, we evaluated two \ac{TA} methods  with eye tracking and identified both as suitable with distinct strengths in different applications.
 Based on two use cases, we demonstrated how eye tracking and \ac{TA}  complement each other for the analysis of \ac{HRE} processes. 
From our results, we derive methodological implications that go beyond the specific domain of \ac{HRE} and propose practical applications in the field of hardware security.

\begin{acks}
% hidden for blind review; also we're not allowed to have names in this document (thanks to MPG privacy policy)
% MPI Python/MatLab student
% MPI qualitative coding student
% WISC eye tracking student? @Jingjie is this correct?
We would like to thank our anonymous reviewers for their constructive feedback on this work.
We are also grateful to Sarah Naqvi, Jannik Schmöle, and Aurelia Sudjana, who were a great help in preparing the user study and in data analysis.

This work was supported by the \grantsponsor{sechuman}{PhD School \enquote{SecHuman -- Security for Humans in Cyberspace} by the federal state of NRW, Germany}{https://sechuman.ruhr-uni-bochum.de/}, and the \grantsponsor{dfg}{German Research Foundation (DFG)}{https://www.dfg.de/en} within the framework of the Excellence Strategy of the Federal Government and the States - \grantnum{dfg}{EXC 2092 CASA - 390781972}.
This work was supported in part by the \grantsponsor{nsf}{U.S. National Science Foundation}{https://www.nsf.gov/} under grants \grantnum{nsf}{1845469}, \grantnum{nsf}{1942014}, and \grantnum{nsf}{2003129}.

We acknowledge the assistance of DALL$\cdot$E 3~\cite{dalle-23} in generating the artistic illustration of Figure~\ref{fig:teaser}. 
\end{acks}

%\paragraph{CRediT authorship contribution statement} We describe each author's contributions with their initials below, following the Contributor Roles Taxonomy (CRediT)~\cite{allen-19}.
\paragraph{\acs{CRediT} authorship contribution statement.} We describe each author's contributions with their initials below, following the \acf{CRediT}~\cite{allen-19}.
\textbf{Conceptualization}: J.L., S.B., C.W., Y.K., K.F., N.R., C.P.
\textbf{Project administration:} R.W., J.L., S.B.
\textbf{Investigation:} J.L., S.B., C.W.
\textbf{Data Curation:} R.W., M.W., J.L.
\textbf{Formal analysis:} R.W., M.W., J.L.
\textbf{Writing -- Original draft:} R.W., M.W., J.L., S.B.
\textbf{Writing -- Review \& Editing:} R.W., M.W., J.L., S.B., M.E., Y.K., K.F., N.R., C.P.
\textbf{Visualization:} R.W., M.W.

\bibliographystyle{ACM-Reference-Format}
\bibliography{bibliography}

\newpage
\appendix
\onecolumn
\section{Full Codebook}
\label{section:appendix:codebook}

Main codes from \autoref{fig:results:rq2:code_freq} are marked in \textbf{bold}.
\bigskip

\noindent\begin{minipage}[t][][t]{0.49\textwidth}
\begin{forest}
  for tree={
    font=\ttfamily,
    grow'=0,
    child anchor=west,
    parent anchor=south,
    anchor=west,
    calign=first,
    s sep=1pt,
    edge path={
      \noexpand\path [draw, \forestoption{edge}]
      (!u.south west) +(7.5pt,0) |- node[fill,inner sep=1.25pt] {} (.child anchor)\forestoption{edge label};
    },
    before typesetting nodes={
      if n=1
        {insert before={[,phantom]}}
        {}
    },
    fit=band,
    before computing xy={l=15pt},
  }
  [Strategies / Reasoning, baseline
   %[consider simulation feedback]
   [\textbf{concluding actions / insight}
    [justification]
    [elimination
     [partial solution]
     [branch]
     [irrelevant input]]
    [decamouflaging]
    [unsuccsessful strategy [perseveres]]
    [reaching a conclusion]]
   [\textbf{theory crafting}
    [planning]
    [theory / assumption / guess [speculation]]
    [validating]]
   [\textbf{backward tracking}]
   [\textbf{forward tracking}]]
\end{forest}

\begin{forest}
  for tree={
    font=\ttfamily,
    grow'=0,
    child anchor=west,
    parent anchor=south,
    anchor=west,
    calign=first,
    s sep=1pt,
    edge path={
      \noexpand\path [draw, \forestoption{edge}]
      (!u.south west) +(7.5pt,0) |- node[fill,inner sep=1.25pt] {} (.child anchor)\forestoption{edge label};
    },
    before typesetting nodes={
      if n=1
        {insert before={[,phantom]}}
        {}
    },
    fit=band,
    before computing xy={l=15pt},
  }
  [\textbf{Exploration}
   [problem exceeds capabilities]
   [starting point identification]
   [localization of camouflaged gate]
   [circuit exploration]
   [goal state identification]]
\end{forest}

\begin{forest}
  for tree={
    font=\ttfamily,
    grow'=0,
    child anchor=west,
    parent anchor=south,
    anchor=west,
    calign=first,
    s sep=1pt,
    edge path={
      \noexpand\path [draw, \forestoption{edge}]
      (!u.south west) +(7.5pt,0) |- node[fill,inner sep=1.25pt] {} (.child anchor)\forestoption{edge label};
    },
    before typesetting nodes={
      if n=1
        {insert before={[,phantom]}}
        {}
    },
    fit=band,
    before computing xy={l=15pt},
  }
  [\textbf{Recall} [recalls prior insight] [recognizes known sub-problem]]
\end{forest}
\end{minipage}%
\begin{minipage}[t][14cm][t]{0.49\textwidth}
\begin{forest}
  for tree={
    font=\ttfamily,
    grow'=0,
    child anchor=west,
    parent anchor=south,
    anchor=west,
    calign=first,
    s sep=1pt,
    edge path={
      \noexpand\path [draw, \forestoption{edge}]
      (!u.south west) +(7.5pt,0) |- node[fill,inner sep=1.25pt] {} (.child anchor)\forestoption{edge label};
    },
    before typesetting nodes={
      if n=1
        {insert before={[,phantom]}}
        {}
    },
    fit=band,
    before computing xy={l=15pt},
  }
  [Errors and Error Correction,baseline
   [\textbf{corrective action}
    [identifying problem]
    [realizing mistake]
    [correcting mistake]]
   [\textbf{error}
    [incorrect reasoning]
    [misinterpretation]
    [forgetting]
    [confusion]
    [input error]]]
\end{forest}

\begin{forest}
  for tree={
    font=\ttfamily,
    grow'=0,
    child anchor=west,
    parent anchor=south,
    anchor=west,
    calign=first,
    s sep=1pt,
    edge path={
      \noexpand\path [draw, \forestoption{edge}]
      (!u.south west) +(7.5pt,0) |- node[fill,inner sep=1.25pt] {} (.child anchor)\forestoption{edge label};
    },
    before typesetting nodes={
      if n=1
        {insert before={[,phantom]}}
        {}
    },
    fit=band,
    before computing xy={l=15pt},
  }
  [Misc
   [\textbf{self-talk} [self instruction]]
   [\textbf{external interruption}]
   [\textbf{participant not talking}]
   [\textbf{mumbles}]]
\end{forest}

\begin{forest}
  for tree={
    font=\ttfamily,
    grow'=0,
    child anchor=west,
    parent anchor=south,
    anchor=west,
    calign=first,
    s sep=1pt,
    edge path={
      \noexpand\path [draw, \forestoption{edge}]
      (!u.south west) +(7.5pt,0) |- node[fill,inner sep=1.25pt] {} (.child anchor)\forestoption{edge label};
    },
    before typesetting nodes={
      if n=1
        {insert before={[,phantom]}}
        {}
    },
    fit=band,
    before computing xy={l=15pt},
  }
  [RTA
   [\textbf{orientation}]
   [\textbf{uncertainty / error}]
   [\textbf{correction}]
   [\textbf{insight}]]
\end{forest}
\end{minipage}

\newpage
\section{Fixation Duration}
\label{section:appendix:fixations}

\begin{figure}[h!]
    \centering
    \includegraphics[width=\linewidth]{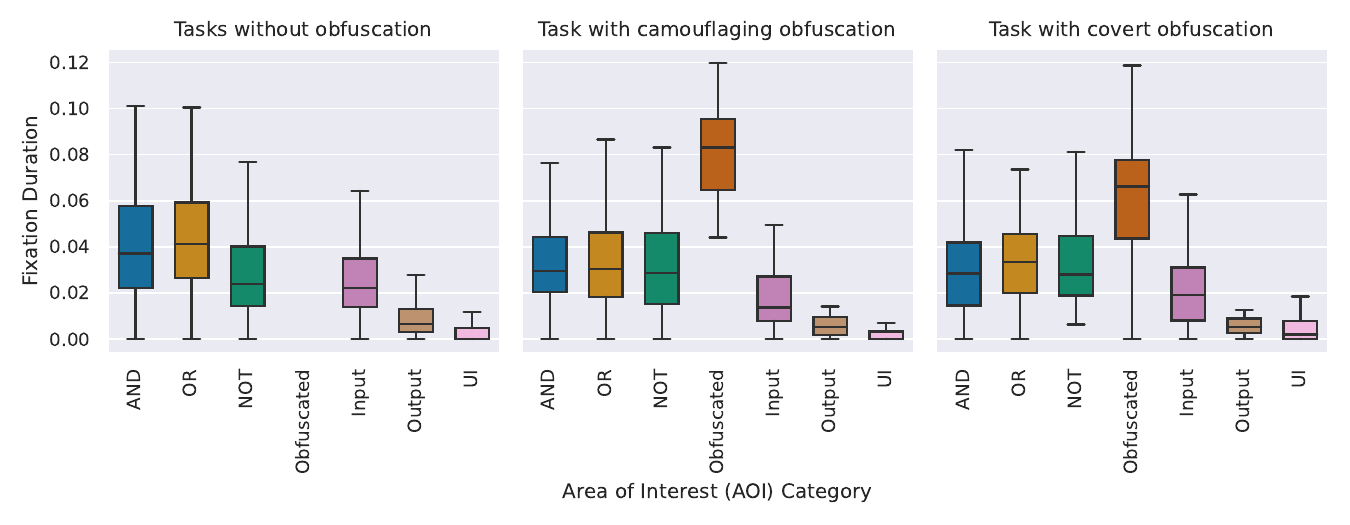}
    \caption{Statistics of fixation duration for each \ac{AOI} category under different task complexities (\ac{RTA} group). The basic logic gate types (AND, OR, NOT) receive similar attention across all types of the \ac{HRE} tasks, but are outweighed by both camouflaged and covert gates. Output and \ac{UI} elements receive little attention.}
    \label{fig:appendix:fixations:gatetype_fixations_time}
    \Description{Figure sixteen is very similar to figure nine. Again, three separate boxplots, one for each HRE task type (regular, camouflaged, covert) are shown. The first plot contains six boxes, the second and third plot contain seven boxes, one box for each AOI type. In this figure, the y-axes display fixation duration instead of fixation rate. Again, it is evident that the obfuscated gate receives the most attention.}
\end{figure}

\end{document}